\documentclass[]{aa}

\usepackage{graphicx}
\usepackage{txfonts}
\usepackage{natbib}

\bibpunct{(}{)}{;}{a}{}{,} 

\newcommand{\herm}{H}
\newcommand{\jones}[2]{\vec {#1}_{#2}}

\newcommand{\jonesT}[2]{\vec {#1}^{\herm}_{#2}}

\newcommand{\coh}[2]{\mathsf{{#1}}_{{#2}}}
\newcommand{\Exp}[1]{\mathrm{e}^{#1}}

\newcommand{\DD}[1]{\,\mathrm{d}{#1}}

\newcommand{\EDIT}[1]{#1}

\begin{document}

\title{Revisiting the radio interferometer measurement equation.\\IV. A generalized tensor formalism}

\author{O.M.\ Smirnov}

\institute{Netherlands Institute for Radio Astronomy (ASTRON)\\
  P.O. Box 2, 7990AA Dwingeloo, The Netherlands \\
  \email{smirnov@astron.nl}}

\date{Received 22 Feb 2011 / Accepted 1 Jun 2011}

\titlerunning{Revisiting the RIME. IV. A generalized tensor formalism}
\authorrunning{O.M.\ Smirnov}

\abstract%
{The radio interferometer measurement equation (RIME), especially in its $2\times2$ form, has provided a \EDIT{comprehensive matrix-based formalism for describing classical radio interferometry and polarimetry}, as shown in the previous three papers of this series. However, recent practical and theoretical developments, such as phased array feeds (PAFs), aperture arrays (AAs) and wide-field polarimetry, are exposing limitations of the formalism.
}
{This paper aims to develop a more general formalism that can be used to both clearly define the limitations of the matrix RIME, and
to describe observational scenarios that lie outside these limitations.
}
{Some assumptions underlying the matrix RIME are explicated and analysed in detail. To this purpose, an array correlation matrix (ACM) formalism is explored. This proves of limited use; it is shown that matrix algebra is simply not a sufficiently flexible tool for the job. To overcome these limitations, a more general formalism based on tensors and the Einstein notation is proposed and explored both theoretically, and with a view to practical implementations.
}%
{The tensor formalism elegantly yields generalized RIMEs describing beamforming, mutual coupling, and wide-field polarimetry in one equation. It is shown that under the explicated assumptions, tensor equations reduce to the $2\times2$ RIME. From a practical point of view, some methods for implementing tensor equations in an optimal way are proposed and analysed.
}%
{The tensor RIME is a powerful means of describing observational scenarios not amenable to the matrix RIME. Even in cases where the latter 
remains applicable, the tensor formalism can be a valuable tool for understanding the limits of such applicability.
}


\maketitle

\section*{Introduction}

Since its formulation by \citet{ME1}, the radio interferometer measurement equation (RIME) has been \EDIT{adopted} by the calibration and imaging algorithm development as the mathematical formalism of choice when describing new methods and techniques for processing radio interferometric data. In its $2\times2$ matrix version (also known as the \emph{Jones formalism\/}, or JF) developed by \citet{ME4}, it has achieved \EDIT{remarkable} simplicity and economy of form. 

Recent developments, however, have begun to expose some limitations of the matrix RIME. In particular, phased array feeds (PAFs) and aperture arrays (AAs), while perfectly amenable to a JF on the systems level (in the sense that the response of a pair of
PAF or AA compound beams can be described by a $2\times2$ Jones matrix), do not seem to fit the same formalism on the element level. In general, since a Jones matrix essentially maps two complex electromagnetic field (EMF) amplitudes onto two feed voltages, it cannot directly describe a system incorporating more than two receptors per station \citep[as in, e.g., the ``tripole'' design of][]{tripole-patent}. And on the flip side of the coin, \citet{Carozzi:ME3D} have shown that two complex EMF amplitudes are insufficient -- even when dealing with only two receptors -- to properly describe wide-field polarimetry, and  that a three-dimensional \emph{Wolf formalism} (WF) is required. Other ``awkward'' effects that don't seem to fit into the JF include mutual coupling of receptors. 

These circumstances seem to suggest that the JF is a special case of some more general formalism, one that is valid only under certain conditions. 
The second part of this paper presents one such generalized formalism. However, given the JF's inherent elegance and simplicity, the degree to which is is understood in the community, and (pragmatically but very importantly) the availability of software implementations, it will in any case 
continue to be a very useful tool. It is therefore important to establish the precise limits of applicability of the JF, which in turn can only be done in the context of a broader theory.

The first part of this paper therefore re-examines the basic tenets of the RIME, and highlights some underlying assumptions that have not been made explicit previously. It then proposes a generalized formalism based on tensors and Einstein notation. As an illustration, some tensor RIMEs are then formulated, for observational scenarios that are not amenable to the JF. The tensor formalism is shown to reduce to the JF under the previously established assumptions. Finally, the paper discusses some practical aspects of implementing such a formalism in software.

\section{Why is the RIME 2$\times$2?}
\label{sec:2x2}

As a starting point, I will consider the RIME formulations derived in Paper~I of this series \citep{RRIME1}. A few crucial equations are reproduced here for reference. Firstly, {\bf the RIME of a point source} gives the visibility matrix measured by interferometer $pq$ as the product of $2\times2$ matrices: the intrinsic source brightness matrix $\coh{B}{}$, and the per-antenna Jones matrices $\jones{J}{p}$ and $\jones{J}{q}$:

\begin{equation}
  \label{eq:rime-ps}
    \coh{V}{pq} = \jones{J}{p} \coh{B}{}  \jonesT{J}{q}.
\end{equation}

The Jones matrix $\jones{J}{p}$ describes the total signal propagation path from source to antenna $p$. For any specific observation and instrument, it is commonly represented by a {\bf Jones chain} of individual propagation effects:

\begin{equation}
  \label{eq:jones-chain}
  \jones{J}{p} = \jones{J}{p,n} \jones{J}{p,n-1} ... \jones{J}{p,1},
\end{equation}

which leads to the {\bf onion form} of the RIME:

\begin{equation}
  \label{eq:rime-onion}
   \coh{V}{pq} = \jones{J}{p,n}(...(\jones{J}{p,2} (\jones{J}{p,1} \coh{B}{}  \jones{J}{q,1}^\herm)\jonesT{J}{q,2}) ... )\jonesT{J}{q,m}
\end{equation}

The individual terms in the matrix product above correspond to different propagation effects along the signal path. Any practical application of the RIME 
requires a set of matrices describing specific effects, which are then inserted into Eq.~(\ref{eq:rime-onion}). These specific matrices tend to have standard single-letter designations \citep[see e.g.][Sect.~7.3]{meqtrees}. In particular, the $K_p$ term\footnote{Following the typographical conventions of Paper~I \citep[][Sect.~1.4]{RRIME1}, I use normal-weight italics for $K_p$ to emphasize the fact that it is a scalar term rather than a full matrix.} describes the geometric (and fringe stopped) phase delay to antenna $p$,  $K_p = \Exp{-2\pi i(u_p l+v_p m+w_p (n-1))}.$ The rest of the Jones chain can be partitioned into direction-independent effects (DIEs, or $uv$-plane effects) on the right, and direction-dependent effects (DDEs, or image-plane effects) on the left, designated as\footnote{Strictly speaking, $\jones{G}{p}$ encompasses the DIEs up to and not including the \emph{leftmost} DDE.} $\jones{G}{p}$ and $\jones{E}{p}$. We can then write a RIME for {\bf multiple discrete sources} as

\begin{equation}
  \label{eq:rime-nps}
  \coh{V}{pq} = \jones{G}{p} \left ( \sum_{s}{\jones{E}{sp} K_{sp} \coh{B}{s} K^\herm_{sq} \jonesT{E}{sq}} \right ) \jonesT{G}{q}.
\end{equation}

Substituting the exponent for the $K_pK_q^\herm$ term then gives us the Fourier transform (FT) kernel in the {\bf full-sky RIME}:

\begin{equation}
  \label{eq:rime-fullsky}
  \coh{V}{pq} = \jones{G}{p} \left( \iint\limits_{lm} \jones{E}{p} \coh{B}{} \jonesT{E}{q} \Exp{-2\pi i(u_{pq} l+v_{pq} m)} \DD{l}\DD{m} \right ) \jonesT{G}{q},
\end{equation}

where all matrix terms\footnote{Note that the $\jones{E}{p}$ term in this equation also incorporates the $w$-term, $W_p=\Exp{-2\pi iw_{pq}n}/\sqrt{n}$,
which allows us to treat the integration as a two-dimensional FT.} under the integration sign are functions of direction $l,m$.

The first fundamental assumption of the RIME is linearity\EDIT{\footnote{Real-life propagation effects are linear either by nature or design, with the exception of a few troublesome regimes, e.g. when using correlators with a low number of bits.}} The second assumption is that the signal is measured in a narrow enough frequency band to be essentially monochromatic, and at short enough timescales that $\jones{J}{p}$ is essentially constant; departures from these assumptions cause \emph{smearing} or \emph{decoherence}, which has already been reviewed in Paper~I \citep[][Sect.~5.2]{RRIME1}. These assumptions are obvious and well-understood. It is more interesting to consider \emph{why\/} the RIME can describe instrumental response by a $2\times2$ Jones matrix. Any such matrix corresponds to a linear transform of two complex number into two complex numbers, so why two and not some other number? This actually rests on some further assumptions.

\subsection{Dual receptors}

\EDIT{In general, an EMF is described by a complex 3-vector $\vec\varepsilon$. However, an EMF propagating as a transverse plane wave can be fully described by only two complex numbers, $\vec e=(e_x,e_y)^T$, corresponding to the first two components of $\vec\varepsilon$ in a coordinate system where the third axis is along the direction of propagation. At the antenna feed, the EMF is converted into two complex voltages $\vec v=(v_a,v_b)^T$. Given a transverse plane wave, two linearly independent complex measurements are necessary and sufficient to fully sample the polarization state of the signal. 

In other words, a $2\times2$ RIME works because we build dual-receptor telescopes; we do the latter because two receptors are what's needed to fully measure the polarization state of a transverse plane wave. PAFs and AAs have more than two receptors, but once these have been electronically combined by a beamformer into a pair of compound beams, any such pair of beams can be considered as a virtual receptor pair for the purposes of the RIME.

\citet{Carozzi:ME3D} have pointed out that in the wide-field case, the EMF arriving from off-axis sources is no longer parallel to the plane of the receptors, so we can no longer measure the polarization state with the same fidelity as for the on-axis case. In the extreme case of a source lying in the plane of the receptors, the loss of polarization information is irrecoverable. Consequently, proper wide-field polarimetry requires three receptors. With only two receptors, the loss-of-fidelity effect can be described by a Jones matrix of its own (which the authors designate as $\jones{T}{}^{(xy)}$), but a fully three-dimensional formalism is required to derive $\jones{T}{}^{(xy)}$ itself.}

\subsection{The closed system assumption}
\label{sec:closed-sys-assumption}

When going from the basic RIME of Eq.~(\ref{eq:rime-ps}) to Eq.~(\ref{eq:rime-onion}), we decompose the total Jones matrix into a chain of propagation effects associated with the signal path from source to station $p$. This is the traditional way of applying the RIME pioneered in the original paper \citep{ME1}, and continued in subsequent literature describing applications of the RIME \citep{JEN:note185,Rau:DDEs,CASA:UserRef,RRIME1}.

Consider an application of Eq.~(\ref{eq:rime-onion}) to real life. Depending on the application, individual components of the Jones chains $\jones{J}{p,i}$ may be derived from {\em a priori} physical considerations and models (e.g. models of the ionosphere), and/or solved for in a closed-loop manner, such as during self-calibration. Crucially, Eq.~(\ref{eq:rime-onion}) postulates that the signal measured by interferometer $pq$ is fully described by the source brightness $\coh{B}{}$ and the set of matrices $\jones{J}{p,i}$ and $\jones{J}{q,j}$, and does not depend on any effect in the signal propagation path to any third antenna $r$. If, however, antenna $r$ is somehow electromagnetically coupled to $p$ and/or $q$, the measured voltages $\vec v_p$ and $\vec v_q$ will contain a contribution received via the signal path to $r$, and thus will have a non-trivial dependence on, e.g., $\jones{J}{r,1}$ that cannot be described by the $2\times2$ formalism alone.

To be absolutely clear, the basic RIME of Eq.~(\ref{eq:rime-ps}) still holds as long as any such coupling is linear. In other words, there is always a single effective $\jones{J}{p}$ that ties the voltage $\vec v_p$ to the source EMF vector $\vec e$. In some applications, e.g. traditional selfcal, where we solve for this $\jones{J}{p}$ in a closed-loop manner, the distinction on whether $\jones{J}{p}$ depends on propagation path $p$ only, or whether other effects are mixed in, is entirely irrelevant. However, when constructing more complicated RIMEs (as is being done currently for simulation of new instruments, or for new calibration techniques), an implicit assumption is made that we may decompose $\jones{J}{p}$ into per-station Jones chains, as in Eq.~(\ref{eq:rime-onion}). This is tantamount to assuming that each station \emph{forms a closed system}.

Consider the effect of electrical cross-talk, or mutual coupling in a densely-packed array. If cross-talk or coupling is restricted to
the two receptors within a station, such a station forms a closed system. For a closed system, the Jones chain approach is perfectly valid. If, however, cross-talk occurs \emph{between receptors associated with different stations}, the receptor voltages $\vec{v}_{p}$ will
not only depend on $\jones{J}{p,i}$, but also on $\jones{J}{q,j}$, $\jones{J}{r,k}$, etc. (See Sect.~\ref{sec:matrix-image-plane} for a more thorough discussion of this point.) With the emergence of AA and PAF designs for new instruments, we can no longer safely assume that two receptors form a closed system; in fact, even traditional interferometers can suffer from troublesome cross-talk in certain situations \citep{ATCA-xtalk}.

Some formulations of the RIME can incorporate coupling within each pair of stations $p$ and $q$ via an additional $4\times4$ matrix \citep[see e.g.][]{JEN:note185} used to describe multiplicative interferometer-based effects. By definition, this approach cannot incorporate coupling with a third station $r$; any such coupling requires additional formulations that are extrinsic to the $2\times2$ RIME, such as the ACM formalism of Sect.~\ref{sec:matrix}, or the tensor formalism that is the main subject of this paper.
 
The closed system assumption has not been made explicit in the literature. This is perhaps due to the fact that the RIME is nominally formulated for a single interferometer $pq$. Consider, however, that for an interferometer array composed of $N$ stations, the ``full'' RIME is actually a set of $N(N-1)/2$ equations. By treating the equations independently, we're implicitly assuming that each equation corresponds to a closed system. The higher-order formalisms derived below will make this issue clear.

\subsection{The colocation assumption}

A final seldom explicated assumption is that each pair of receptors is \emph{colocated}. \EDIT{While not required for the general RIME formulation of Eq.~(\ref{eq:rime-ps}) per se, colocation becomes important (and is quietly assumed) in specific applications for two reasons.} Firstly, it allows us to consider the geometric phase delay of both receptors to be the same, which makes the $K_p$ matrix scalar, and allows us to commute it around the Jones chain. $K_p$ and $K_q^\herm$ can then be commuted together to form the FT kernel, which is essential for deriving the full-sky variants of the RIME such as Eq.~(\ref{eq:rime-fullsky}). And secondly, although the basic RIME of Eq.~(\ref{eq:rime-ps}) may be formulated for any four arbitrarily-located receptors, when we proceed to decompose $\jones{J}{p}$ into per-station terms, we implicitly assume a single propagation path per each pair of receptors (same atmosphere, etc.), which implies colocation. In practice the second consideration may be negligible, but not so the first.

Classical single-feed dish designs have colocated receptors as a matter of course, but a PAF system such as APERTIF \citep{Apertif} or ASKAP \citep{ASKAP} typically has horizontally and vertically oriented dipoles at slightly different positions. The effective phase centres of the beamformed signals may be different yet again. The $\jones{K}{p}$ matrix then becomes diagonal but not scalar, and can no longer be commuted around the RIME. In principle, we can shoehorn the case of non-colocated receptors into the RIME formulations by picking a reference point (e.g., the  mid-point between the two receptors), and decomposing $\jones{K}{p}$ into a product of a scalar phase delay corresponding to the reference point, and a non-scalar differential delay term: $\jones{K}{p}=K_p^{(0)}\jones{K}{p}^{(\delta)}.$ The scalar term $K_p^{(0)}$ can then be commuted around the RIME to yield the FT kernel of Eq.~(\ref{eq:rime-fullsky}), while $\jones{K}{p}^{(\delta)}$ becomes a DIE that can be absorbed into the overall phase calibration (or cause instrumental $V$ or $U$ polarization if it isn't). The exact form of $\jones{K}{p}^{(\delta)}$ and $K_p^{(0)}$ can be derived from geometric considerations (or analysis of instrument optics), but such a derivation is extrinsic to the RIME per se. This situation is similar to that of the $\jones{T}{}^{(xy)}$ term derived by \citet{Carozzi:ME3D}, and is another reason behind the multidimensional formalism proposed later on in this paper.

Note that conventional FT-based imaging algorithms also assume colocated receptors when converting visibilities to Stokes parameters. For example, the conventional formulae for $I$ and $U$,

\[
I=\frac{1}{2}(V_{xx}+V_{yy}),\;\;U=\frac{1}{2}(V_{xy}+V_{yx}),
\] 

implicitly assume that the constituent visibilities are measured on the same baseline. Some leeway is acceptable here: since the measured visibilities are additionally convolved by the aperture illumination function (AIF), the formulae above still apply, as long as the degree of non-colocation is negligible compared to the effective station size. Note also that some of the novel approaches of expectation-maximization imaging \citep{leshem-em,levanda-leshem-em} formulate the imaging problem in such a way that the colocation requirement can probably be done away with altogether.

\section{The array correlation matrix formalism}
\label{sec:matrix}

I will first explore the limitations of the closed-system assumption a little bit further. Consider an AA, PAF, or conventional closely-packed interferometer array where mutual coupling affects more than two receptors at a time. Such an array cannot be partitioned into pairs of receptors, with each pair forming a closed system. The normal $2\times2$ RIME of Eq.~(\ref{eq:rime-onion}) is then no longer valid. An alternative is to describe the response of such an array in terms of a single array correlation matrix (ACM, also called the signal voltage covariance matrix), as has been done by \citet{wijnholds-thesis} for AAs, and \citet{warnick-paf-polarimetry} for PAFs. Since the ACM provides a valuable conceptual link between the $2\times2$ RIME and the tensor formalism described later in this paper, I will consider it in some detail in this section.

Let's assume an arbitrary set of $n$ receptors (e.g. dipoles) in arbitrary orientation, and a single point source of radiation. 
If we represent the voltage response of the full array by the column vector $\vec{v}=(v_{1}\ldots v_{n})^T$, then we can express it (assuming linearity as usual) as the product of a $n\times2$ matrix with the source EMF vector $\vec e$:

\[
\vec{v} = \mathbf{J}\vec e = \left(\begin{array}{cc}
j_{x1} & j_{y1}\\
\vdots & \vdots\\
j_{xn} & j_{yn}\end{array}\right)
\left(\begin{array}{c} e_x\\e_y
\end{array}\right)
\]

If all pairwise combinations of receptors are correlated, we end up with an $n\times n$ ACM\footnote{I will use boldface roman capitals, such as $\mathbf{V}$, for matrices other than $2\times2$. As per the conventions of Paper~I \citep[][Sect.~1.4]{RRIME1}, $2\times2$ Jones matrices are indicated by boldface italic capitals ($\jones{J}{}$). Brightness, coherency and visibility matrices (as well as tensors, introduced later) are indicated by sans-serif capitals, e.g. $\coh{B}{}$.} $\mathbf{V}$:

\begin{equation}\label{eq:array-response-j}
\mathbf{V} =\langle \vec v \vec v^\herm \rangle =\left(\begin{array}{cc}
j_{x1} & j_{y1}\\
\vdots & \vdots\\
j_{xn} & j_{yn}\end{array}\right) \langle \vec e \vec e^\herm \rangle \left(\begin{array}{ccc}
j^*_{x1} & \cdots & j^*_{xn}\\
j^*_{y1} & \cdots & j^*_{yn}\end{array}\right)= \mathbf{J} \coh{B}{} \mathbf{J}^\herm,
\end{equation}

where $\coh{B}{}$ is the $2\times2$ source brightness matrix, and $\mathbf{J}$ is an $n\times2$ Jones-like matrix for the entire array. Note that for $n=2$, this equation becomes the autocorrelation matrix given by the RIME of Eq.~(\ref{eq:rime-ps}) with $p=q$.

To derive the $\mathbf{J}$ matrix for a given observation, we need to decompose it into a product of ``physical'' terms that we can
analyse individually. As an example, let's consider only three effects: primary beam (PB) gain, geometric phase, and cross-talk. The 
$\mathbf{J}$ matrix can then be decomposed as follows:

\begin{equation}\label{eq:array-response}
\vec v = \mathbf{QKE} \vec e =\left(\begin{array}{ccc}
q_{11} & \ldots & q_{1n}\\
 & \ddots\\
q_{n1} & \ldots & q_{nn}\end{array}\right)\left(\begin{array}{ccc}
\kappa_{1} &  & 0\\
 & \ddots & \\
0 &  & \kappa_{n}\end{array}\right)\left(\begin{array}{cc}
\epsilon_{x1} & \epsilon_{y1}\\
\vdots & \vdots\\
\epsilon_{xn} & \epsilon_{yn}\end{array}\right) 
\left(\begin{array}{c} e_x\\e_y
\end{array}\right)
\end{equation}

and the full ME then becomes:

\begin{equation}
\mathbf{V} = \mathbf{QKE} \coh{B}{} \mathbf{E}^\herm \mathbf{K}^\herm \mathbf{Q}^\herm.
\label{eq:array-me}
\end{equation}

The $n\times2$ \textbf{$\mathbf{E}$} matrix corresponds to the PB gain,
the $n\times n$ diagonal $\mathbf{K}$ matrix corresponds to the individual
phase terms (different for every receptor), and the $n\times n$ $\mathbf{Q}$
matrix corresponds to the cross-talk and/or mutual coupling between the
receptors. The equation does not include an explicit term for the 
complex receiver gains: these can be described either by a separate diagonal matrix,
or absorbed into $\mathbf{Q}$.

In the case of a classical array of dishes, we have $n=2m$ receptors, with each
adjacent pair forming a closed system. In this case, $\mathbf{Q}$ becomes block-diagonal -- that is, composed of $2\times2$ blocks along the diagonal, equivalent to the ``leakage'' matrices of the original RIME formulation \citep{ME1}. $\mathbf{K}$ becomes block-scalar ($\kappa_{1}=\kappa_{2},\kappa_{3}=\kappa_{4},...)$, and Eq.~(\ref{eq:array-me}) 
dissolves into the familiar set of $m(m-1)/2$ independent RIMEs of Eq.~(\ref{eq:rime-onion}).

Note that the ordering of terms in this equation is not entirely physical
-- in the actual signal path, the phase delay represented by $\mathbf{K}$
occurs before the beam response $\mathbf{E}$. To be even more precise, phase 
delay may be a combination of geometric phase that occurs ``in space'' before $\mathbf{E}$, and fringe stopping 
that occurs ``in the correlator'' after $\mathbf{Q}$. Such an ordering of effects 
becomes very awkward to describe with this matrix formalism,  
but will be fully addressed by the tensor formalism of Sect.~\ref{sec:tensor}.

\subsection{Image-plane effects and cross-talk}
\label{sec:matrix-image-plane}

If we now consider additional image-plane effects\footnote{The previous papers in this series \citep{RRIME1,RRIME2,RRIME3} also refer to these as direction-dependent effects (DDEs). As far as the present paper is concerned, the important aspect of these effects is that they arise \emph{before} the receptor, rather than their directional-dependence per se. I will therefore use the alternative term \emph{image-plane effects} in order to emphasize this.}, things get even more awkward. In the simple case,
if these effects do not vary over the array (i.e. for a given direction, are the same along each line of sight to each receptor), 
we can replace the $\vec e$ vector in Eq.~(\ref{eq:array-response}) by $\jones{Z}{}\vec e$, where $\jones{Z}{}$ is a Jones matrix
describing the image-plane effect. We can then combine the $n\times2$
\textbf{E} matrix and the $2\times2$ $\jones{Z}{}$ matrix into a single
term $\mathbf{R}=\mathbf{E}\jones{Z}{}$, which is a $n\times2$ matrix describing
the voltage gain and all other image-plane effects, and define an $n\times n$ ``apparent
sky'' matrix as $\mathbf{B}_\mathrm{app} = \mathbf{R} \coh{B}{} \mathbf{R}^\herm$.  
Equation~(\ref{eq:array-me}) then becomes

\[
\mathbf{V}=\mathbf{QKB}_\mathrm{app} \mathbf{K}^\herm \mathbf{Q}^\herm.
\]

If image-plane effects do vary across receptors, then a matrix formalism
is no longer sufficient! The expression for each receptor
$p$ must somehow incorporate its own $\jones{Z}{p}$ Jones matrix.
We need to describe signal propagation along $n$ lines of sight,
and each propagation effect needs a $2\times2$ matrix. A full description of 
the image-plane term then needs $n\times2\times2$ complex numbers. 

Another way to look at this conundrum is as follows. As long as each
receptor pair is colocated and forms a closed system (as is the case
for traditional interferometers), the voltage response of each receptor
depends only on the EMF vector at its location. The correlations between
stations $p$ and $r$ can then be fully described in terms of the
EMF vectors at locations $p$ and $r$. This allows us to write the RIME
in a matrix form, as in Eq.~(\ref{eq:rime-onion}) or (\ref{eq:array-me}). In the presence of significant
cross-talk between more than two receptors, the voltage response of
each receptor depends on the EMF vectors at multiple locations. In effect,
the cross-talk term $\mathbf{Q}$ in Eq.~(\ref{eq:array-me}) ``scrambles up'' image plane effects between
different receptor locations; describing this is beyond the capability of ordinary matrix algebra.

In practice, receptors that are sufficiently separated to see any conceivable difference in image-plane effects would be too 
far apart for any mutual coupling, while today's all-digital designs have also eliminated most possibilities of 
cross-talk. Mathematically, this corresponds to $\jones{Z}{p} \approx \jones{Z}{r}$ where $q_{pr} \ne 0$, which means that image-plane 
effects can, in principle, be shoehorned into the matrix formalism of Eq.~(\ref{eq:array-me}). This, however, 
does not make the formalism any less clumsy -- we still need to describe different image-plane effects for far-apart receptors, and mutual coupling for close-together ones, and the two effects together are difficult to shoehorn into ordinary matrix multiplication. 

\subsection{The non-paraxial case}
\label{sec:matrix-3d}

\citet{Carozzi:ME3D} have shown that the EMF can only be accurately described by a 2-vector in the paraxial or nearly-paraxial case. For wide-field polarimetry, we must describe the EMF by a rank-3 column vector $\vec\varepsilon=(e_x,e_y,e_z)^T$, and the sky brightness distribution by a $3\times3$ matrix $\mathbf{B}^{(3)}=\langle \vec\varepsilon \vec\varepsilon^\herm \rangle.$ The intrinsic sky brightness is still given by a $2\times2$ matrix $\coh{B}{}$; once an $xyz$ Cartesian system is established, this maps to $\mathbf{B}{}^{(3)}$ via a $3\times2$ transformation matrix $\mathbf{T}$ ({\em ibid.}, Eqs.~(20) and (21)):

\[
  \mathbf{B}^{(3)} = \mathbf{T} \coh{B}{} \mathbf{T}^T.
\]

It is straightforward to incorporate this into the ACM formalism: the $\coh{B}{}$ term of Eqs.~(\ref{eq:array-me}) 
is replaced by $\mathbf{B}^{(3)}$, and the dimensions of the $\mathbf{E}$ matrix become $n\times3$.

\section{A tensor formalism for the RIME}
\label{sec:tensor}

The ACM formalism of the previous section turns out to be only marginally useful for the purposes of this paper. It does aid in understanding the effect of mutual coupling and the closed system assumption a little bit better, but it is much too clumsy in describing image-plane effects, principally because the rules of matrix multiplication are too rigid to represent this particular kind of linear transform. What we need is a more flexible scheme for describing arbitrary multi-linear transforms, one that can go beyond vectors and matrices. Fortunately, mathematicians have already developed just such an apparatus in the form of \emph{tensor algebra}. In this section, I will apply this to derive a generalized multi-dimensional RIME.

\subsection{Tensors and the Einstein notation: a primer}

Tensors are a large and sprawling subject, and one not particularly familiar to radio astronomers at large. Appendix~\ref{sec:tensor-theory} provides a brief but formal description of the concepts required for the formulations of this paper. This is intended for the in-depth reader (and to provide rigorous mathematical underpinnings for what follows). For an executive overview, only a few basic concepts are sufficient:
  
\noindent{\bf Tensors} are a generalization of vectors and matrices. An \emph{$(n,m)$-type} tensor is given by an $(n+m)$-dimensional array of numbers, and written using $n$ upper and $m$ lower indices: e.g. $\tens{T}_{j_{1}j_{2}...j_{m}}^{i_{1}i_{2}...i_{n}}$. Superscripts are indices just like subscripts, and not exponentiation\footnote{This is the way they will be used in this paper from this point on, with the exception of small integer powers (e.g. $l^2$), where exponentiation is obviously implied.}! For example, a vector is typically a (1,0)-type tensor, denoted as $x^i$. A matrix is a (1,1)-type tensor, denoted as $\tens{A}^i_j$.

\EDIT{\noindent{\bf Upper and lower} tensor indices are quite distinct, in that they determine how the components of a tensor behave under coordinate transforms. Upper indices are called {\em contravariant}, since components with an upper index (such as the components of a vector $x_i$) transform reciprocally to the coordinate frames. As a simple example, consider a ``new'' coordinate frame whose basis vectors are scaled up by a factor of $a$ with respect to those of the ``old'' frame. In the ``new'' frame, the same vector is then described by coordinate components that are scaled by a factor of $a^{-1}$ w.r.t. the ``old'' components. By contrast, for a {\em linear form} $f_j$ (that is, a linear function mapping vectors to scalars), the ``new'' components are scaled by a factor of $a$. Lower indices are thus said to be {\em covariant}. 

In physical terms, upper indices tend to refer to vectors, and lower indices to linear functions on vectors. An $n\times n$ matrix can be thought of as a ``vector'' of $n$ linear functions on vectors, and thus has one upper and one lower index in tensor notation, and transforms both co- and contravariantly. This is manifest in the familiar $\mathbf{T}^{-1}\mathbf{AT}$ (or $\mathbf{T}\mathbf{A}\mathbf{T}^{-1}$, depending which way the coordinate transform matrix $\mathbf{T}$ is defined) formula for matrix coordinate transforms. For higher-ranked tensors, the general rules for coordinate transforms are covered in Sect.~\ref{sec:coordinatre-xforms}.}
 
\noindent{\bf Einstein notation} (or Einstein summation) is a convention whereby repeated upper and lower indices in a product of tensors are implicitly summed over. For example, 

\[
  y^i = \tens{A}^i_j x^j = \sum_{j=1}^{N} \tens{A}^i_j x^j
\]

is a way to write the matrix/vector product $\vec y=\mathbf{A} \vec x$ in Einstein notation. The $j$ index is a \emph{summation} index since it is repeated, and the $i$ index is a \emph{free} index. Another useful convention is to use Greek letters for the summation indices. For example, a matrix product may be written as $\tens{C}^i_j = \tens{A}^i_\alpha \tens{B}^\alpha_j$.

\noindent{\bf The tensor conjugate}  is a generalization of the Hermitian transpose. This is indicated by a bar over the symbol and a swapping of the upper and lower indices. For example, $\bar x_i$ is the conjugate of $x^i$, and $\bar\tens{A}^i_j$ is the conjugate of $\tens{A}^j_i$.

\subsection{Recasting the RIME in tensor notation}
\label{sec:tensor-rime}

As an exercise, let's recast the basic RIME of Eq.~(\ref{eq:rime-ps}) using tensor notation. This essentially repeats the derivations 
of Paper~I \citep{RRIME1} using tensor terminology (compare to Sect.~1 therein). 

For starters, we must define the underlying vector space. The classical Jones formalism (JF) corresponds to rank-2 vectors, i.e. the $\mathbb{C}^2$ space.
We can also use $\mathbb{C}^3$ space instead, which results in a version of the Wolf formalism (WF) suggested by \citet{Carozzi:ME3D}. Remarkably, both
formulations look exactly the same in tensor notation, the only difference being the implicit range of the tensor indices. I'll stick to the familiar  terminology of the JF here, but the same analysis applies to the WF.

An EMF vector is then just a (1,0)-type tensor $e^i$. Linear transforms of vectors (i.e. Jones matrices) correspond to (1,1)-type tensors, $[\tens{J_p}]_j^i$ (note that $\tens{p}$ is not, as yet, a tensor index here, but simply a station ``label'', which is emphasized by hiding it within brackets). The voltage response of station $p$ is then

\[
[v_\tens{p}]^i=[\tens{J_p}]_\alpha^i e^\alpha,
\]

where $\alpha$ is a summation index. The coherency of two voltage or EMF vectors is defined via the outer product\footnote{This definition is actually fraught with subtleties: see Sect.~\ref{sec:coh-outer-prod} for a discussion.} $e^i \bar e_j$, yielding a (1,1)-type tensor, i.e. a matrix: 

\[
[\tens{V_{pq}}]^i_j = 2 \langle [v_\tens{p}]^i [\bar v_\tens{q}]_j \rangle,
\]

Combining the two equations above gives us

\[
[\tens{V_{pq}}]^i_j = 2 
\left \langle 
  \left[ [\tens{J_p}]_{\alpha}^{i} e^{\alpha} \right] 
  \left[ [\tens{J_q}]_{\beta}^{j}e^{\beta}    \right]^* 
\right \rangle =
[\tens{J_p}]_{\alpha}^{i}(2 \langle e^\alpha\bar e_\beta \rangle)[\bar{\tens{J_q}}]_{j}^{\beta}
\]

And now, defining the source brightness tensor $\tens{B}^i_j$ as $2\langle e^i \bar e_j \rangle$, we arrive at

\begin{equation}
[\tens{V_{pq}}]^i_j = 
[\tens{J_p}]_{\alpha}^{i} \tens{B}^\alpha_\beta [\bar\tens{J_q}]_{j}^{\beta},
\label{eq:me0-tensor}
\end{equation}

which is exactly the RIME of Eq.~(\ref{eq:rime-ps}), rewritten using Einstein notation. Not surprisingly, it looks somewhat more bulky than the original -- matrix multiplication, after all, is a more compact notation for this particular operation.

Now, since we can commute the terms in an Einstein sum (as long as they take their indices with them, see Sect.~\ref{sec:einstein-commutation}), we can split off the two $\tens{J}$ terms into a sub-product which we'll designate as $[\tens{J_{pq}}]$:

\begin{equation}
\label{eq:me0-tensor-Jpq}
[\tens{J_p}]_{\alpha}^{i} \tens{B}^\alpha_\beta [\bar\tens{J_q}]_{j}^{\beta} = 
\left( [\tens{J_p}]_{\alpha}^{i}  [\bar\tens{J_q}]_{j}^{\beta} \right ) \tens{B}^\alpha_\beta =
[\tens{J_{pq}}]_{j \alpha}^{i \beta} \tens{B}^\alpha_\beta.
\end{equation}

What is this $\tens{J_{pq}}$? It is a (2,2)-type tensor, corresponding to $2\times2\times2\times2=16$ numbers. Mathematically, it is the exact equivalent of the outer product $\jones{J}{p}\otimes \jonesT{J}{q}$,  giving us the $4\times4$ form of the RIME, as originally formulated by \citet{ME1}. The components of the tensor given by $[\tens{J_{pq}}]_{j \alpha}^{i \beta} \tens{B}^\alpha_\beta$ correspond exactly to the components of a 4-vector produced via multiplication of the $4\times4$ matrix $\jones{J}{p}\otimes \jonesT{J}{q}$ by the 4-vector $\mathbf{S}\vec I$ \citep[see Paper~I,][Sect.~6.1]{RRIME1}.

Finally, note that we've been ``hiding'' the $p$ and $q$ station labels inside square brackets, since they 
don't take part in any tensor operations above. Upon further consideration, this distinction proves to be 
somewhat artificial. Let's treat $p$ and $q$ as \emph{free} tensor indices in their own right\footnote{Strictly speaking, all tensor
indices should have the same range. I'm implicitly invoking \emph{mixed-dimension tensors} at this point. See Sect.~\ref{sec:tensors-mixed-dim} for a formal treatment of this issue.}. The set of all
Jones matrices for the array can then be represented by a (2,1)-type tensor $\tens{J}^{pi}_j.$ All the visibilities measured by the array as a whole
will then be represented by a (2,2)-type tensor $\tens{V}^{pi}_{qj}$, and we can then rewrite Eq.~(\ref{eq:me0-tensor}) as

\begin{equation}
\tens{V}^{pi}_{qj}=\tens{J}_{\alpha}^{pi} \tens{B}^\alpha_\beta \bar\tens{J}^{\beta}_{qj},
\label{eq:me0-tensor1}
\end{equation}

...which is now a single equation for all the visibilities measured by the array, \emph{en masse} (as opposed to the visibility of a single baseline given by Eq.~(\ref{eq:rime-ps}) or (\ref{eq:me0-tensor})). Such manipulation of tensor indices may seem like a purely formal trick, but will in fact prove very useful when we consider generalized RIMEs below.

Note that the brightness tensor $\tens{B}$ is self-conjugate (or Hermitian), in the sense that $\tens{B}^i_j=\bar\tens{B}^j_i$. The visibility tensor $\tens{V}$, on the other hand, is only Hermitian with respect to a permutation of $p$ and $q$: $\tens{V}^{pi}_{qj}=\overline{ \tens{V}^{qi}_{pj}}.$ 

\section{Generalizing the RIME}

In this section, I will put tensor notation to work to incorporate image-plane effects and mutual coupling and beamforming into a generalized RIME hinted at in  Sect.~\ref{sec:matrix}. This shows how the formalism may be used to derive a few  different forms of the RIME for various instrumental scenarios. Note that the resulting equations are somewhat speculative, and not necessarily applicable to any particular real-life instrument. The point of the exercise is to demonstrate the flexibility of the formalism in deriving RIMEs that go beyond the capability of the Jones formalism. 

First, let's set up some indexing conventions. I'll use $i,j,k,...$ for free indices that run from 1 to $N=2$ (or 3, see below), i.e. for those that refer to EMF components, or voltages on paired receptors, and $\alpha,\beta,\gamma,...$ for summation indices in the same range. I shall refer to such indices as \emph{2-indices} (or 3-indices). For free indices that refer to stations or disparate receptors (and run from 1 to $N$), I'll use $p,q,r,s,...$, and for the corresponding summation indices, $\sigma,\tau,\upsilon,\phi,...$ I shall refer to these as \emph{station indices}.

Consider again the $N$ arbitrary receptors of Sect.~\ref{sec:matrix} observing a single source. The source EMF is given by the tensor $e^i.$ 
All effects between the source and the receptor, up to and not including the voltage gain, can be described by a (2,1)-type tensor, $\tens{Z}_{j}^{pi}.$ This implies that they are different for each receptor $p$. The PB response of the  receptor can be described by a (1,1)-type tensor, $\tens{E}_i^p$.
Finally, the geometric phase delay of each receptor is a (1,0)-type tensor, $\tens{K}^{p}.$

Let's take this in small steps. The EMF field arriving at each receptor $p$ is given by

\begin{equation}
\label{eq:gme-deriv-KZe}
e'^{pi}=\tens{K}^{p}\tens{Z}_{\alpha}^{pi}e^{\alpha}
\end{equation}

(remembering that we implicitly sum over $\alpha$ here). If we consider just one receptor in isolation, we can re-write the equation for one specific value of $p$. This corresponds to  the familiar matrix/vector product:

\[
e'^i = \tens{K}\tens{Z}_{\alpha}^{i}e^{\alpha},\;\;\;\mbox{or}\;\;\;\vec e' = K \jones{Z}{} \vec e,
\]

where $\jones{Z}{}$ is the Jones matrix describing the image-plane effect for this particular receptor, 
and $K$ is the geometric phase delay. The receptor translates the EMF vector $e'^{i}$ into a scalar voltage $v'$.  
This is done via its PB response tensor, $\tens{E}_i$, which is just a row vector:

\[
v'=\tens{E}_{\beta}e'^{\beta}.
\]

Now, if we put the receptor index $p$ back in the equations, we arrive at the tensor expression:

\begin{equation}
\label{eq:gme-deriv-EKZe}
v'^{p}=\tens{E}_{\beta}^{p}\tens{K}^{p}\tens{Z}_{\alpha}^{p\beta}e^{\alpha}
\end{equation}

We're now summing over $\alpha$ when applying image-plane effects, and over $\beta$ when applying the PB response.

Finally, cross-talk and/or mutual coupling scrambles the receptor voltages. If $v'^{p}$
is the ``ideal'' voltage vector without cross-talk, then we need to multiply it by an $n\times n$ matrix (i.e. a (1,1)-type tensor)
to apply cross-talk:

\begin{equation}
\label{eq:gme-deriv-Qv}
v^{p}=\tens{Q}_{\sigma}^{p}v'^{\sigma}.
\end{equation}

The final equation for the voltage response of the array is then:

\begin{equation}
v^{p}=\tens{Q}_{\sigma}^{p}\tens{E}_{\beta}^{\sigma}\tens{K}^{\sigma}\tens{Z}_{\alpha}^{\sigma\beta} e^{\alpha}.
\label{eq:v-tensor}
\end{equation}

We're now summing over $\sigma$ (which ranges over all receptors), $\alpha$ and $\beta$. 

The visibility tensor $\tens{V}^p_q$, containing all the pairwise correlations between the receptors, can then be computed as 
$2\langle v^{p}\bar{v}_{q} \rangle$. Applying Eq.~(\ref{eq:v-tensor}), this becomes

\[
\tens{V}^p_q = 2 \left\langle 
  \left[ \tens{Q}_{\sigma}^{p}\tens{E}_{\beta}^{\sigma}\tens{K}^{\sigma}\tens{Z}_{\alpha}^{\sigma\beta} e^{\alpha} \right] \,
  \left[ \tens{Q}_{\tau}^{q}\tens{E}_{\gamma}^{\tau}\tens{K}^{\tau}\tens{Z}_{\delta}^{\tau\gamma} e^{\delta} \right]^{*} \right\rangle
\]

This uses a different set of summation indices within each pair of brackets, since each sum is computed independently.
Doing the conjugation and rearranging the terms around, we arrive at:

\begin{equation}
\tens{V}^p_q = 
  \tens{Q}_{\sigma}^{p}\tens{E}_{\beta}^{\sigma}\tens{K}^{\sigma}\tens{Z}_{\alpha}^{\sigma\beta}
  \tens{B}^\alpha_\delta
  \bar\tens{Z}_{\tau\gamma}^{\delta}\bar\tens{K}_{\tau}\bar\tens{E}_{\tau}^{\gamma}\bar\tens{Q}_{q}^{\tau}.
\label{eq:gme0}
\end{equation}

This is the tensor form of a RIME for our hypothetical array. Note that structurally it is quite similar 
to the ACM form of Sect.~\ref{sec:matrix} (e.g. Eq.~(\ref{eq:array-me})), but with one principal difference: the (2-1)-type $\tens{Z}$ tensor describes receptor-specific effects, which cannot be expressed via a matrix multiplication. Note also that the other awkwardness encountered in Sect.~\ref{sec:matrix}, namely the difficulty of putting geometric phase delay and fringe stopping into their proper places in the equation, is also elegantly addressed by the tensor formalism. Additional phase delays tensors can be inserted at any point of the equation.

\subsection{Wolf vs. Jones formalisms}

Equation~(\ref{eq:gme0}) generalizes both the classical Jones formalism (JF), and the three-component Wolf formalism (WF).
The JF is constructed on top of a two-dimensional vector space: EMF vectors have two components, the indices 
$\alpha,\beta,...$ range from 1 to 2, and the $\tens{B}$ tensor is the usual $2\times2$ brightness matrix. The WF 
corresponds to a three-dimensional vector space, with the $\tens{B}$ tensor becoming a $3\times3$ matrix. 

Recall (Sect.~\ref{sec:matrix-3d}) that the $3\times3$ brightness matrix is derived from the $2\times2$ brightness matrix via a $3\times2$ transformation matrix $\mathbf{T}$. This derivation can also be expressed as an Einstein sum:

\[
  [\tens{B^{(3)}}]^i_j = \tens{T}^i_{\alpha} [\tens{B^{(2)}}]^\alpha_\beta \tens{T}_j^\beta,
\]
where $\tens{T}$ is the tensor equivalent of the transformation matrix.

In subsequent formulations, I will make no distinction between the JF and the WF unless necessary, with the implicit understanding that the appropriate indices range from 1 to 2 or 3, depending on which version of the formalism is needed.

\subsection{Decomposing the $\tens{J}$ matrix}

If we isolate the left-hand sub-product in Eq.~(\ref{eq:gme0}),

\[
  \tens{Q}_{\sigma}^{p}\tens{E}_{\beta}^{\sigma}\tens{K}^{\sigma}\tens{Z}_{\alpha}^{\sigma\beta},
\]

and track down the free indices in this tensor expression -- $p$ and $\alpha$ -- we can see that the product is a (1,1) tensor, 
$\tens{J}_{\alpha}^{p}.$ We can then rewrite the equation in a more compact form:

\begin{equation}
\tens{V}^p_q = 
  \tens{J}_\alpha^p
  \tens{B}^\alpha_\beta
  \bar\tens{J}^\beta_q
\label{eq:gme-J}
\end{equation}

Not surprisingly, this is just the ACM RIME of Eq.~(\ref{eq:array-response-j}) rewritten in Einstein notation.
In hindsight, this shows how we can break down the full-array response matrix $\mathbf{J}$ into a tensor product of physically meaningful
terms. Note how this parallels the situation of the $2\times 2$ form RIME: even though each $2\times2$ visibility matrix, in principle, depends on 
only two Jones matrices (Eq.~(\ref{eq:rime-ps})), in real-life applications we almost always need to form them up from a chain  of several different 
Jones terms, as in e.g. the onion form (Eq.~(\ref{eq:rime-onion})). What the tensor formulation offers is simply a more capable means of computing the response matrices (more capable than a matrix product, that is) from individual \emph{propagation tensors}.

\subsection{Characterizing propagation tensors}

Since it the original formulation of the matrix RIME, a number of standard types of Jones matrices have seen widespread use.
The construction of Jones matrices actually follows fairly simple rules (even if their behaviour as a function of time, frequency and direction may be quite complicated). A number of similar rules may be proposed for propagation tensors:

\begin{itemize}
\item A tensor that translates the EMF vector into another vector (e.g., Faraday rotation) must necessarily have an upper and a lower
2-index. 
\item A tensor that translates both components of the EMF field equally (i.e. a scalar operation such as phase delay) does not need any 
-indices at all.
\item A tensor transforming the EMF vector into a scalar (e.g., the voltage response of a receptor) must have a lower 2-index.
\item A tensor for an effect that is different across receptors must have a station index.
\item A tensor for an effect that maps per-receptor quantities onto per-receptor quantities must
have two station indices (upper and lower).
\end{itemize}

Some examples of applying these rules:

\begin{itemize}
\item Faraday rotation translates vectors, so it must have an upper and
a lower 2-index. If different across stations and/or receptors, it must also
have a station index. This suggests that the tensor looks like $\tens{F}_{j}^{pi}$ (or $\tens{F}_{pj}^{i}$).

\item Phase delay operates on the EMF vector as a scalar. It is different
across receptors, hence its tensor looks like $\tens{K}^{p}$.
\item PB response translates the EMF vector into a scalar voltage, and
must therefore have one lower 2-index. It is usually different across stations and/or receptors, hence its tensor looks 
like $\tens{E}_{i}^{p}.$
\item Cross-talk or mutual coupling translates receptor voltages into receptor voltages, so it needs two station indices. Its tensor looks
like $\tens{Q}_{q}^{p}$.
\item If mutual coupling needs to be expressed in terms of the EMF field at each receptor instead,
then it may need two 2-indices and two station indices, giving a (2,2)-type tensor, $\tens{Q}_{qj}^{pi}$.
Alternatively, this may be combined with the PB response tensor $\tens{E}$, giving the voltage response of each receptor as
a function of the EMF vector at all the other receptors. This would be a (2,1)-tensor, $\tens{E}_{j}^{pi}$.
\end{itemize}

\subsection{Describing mutual coupling}

Equations (\ref{eq:v-tensor}) and (\ref{eq:gme0}) were derived under the perhaps simplistic assumption that the effect\footnote{As opposed
to the mechanism, which is considerably more complex, and outside the scope of this work.} of mutual coupling can be fully described via cross-talk between the receptor voltages. That is, the collection of EMF vectors at receptor's location was described by a (2,0)-type tensor, $e^{pi}$ (Eq.~(\ref{eq:gme-deriv-KZe})), then converted into nominal receptor voltages by the PB tensor $\tens{E}^p_i$ (Eq.~(\ref{eq:gme-deriv-EKZe})), and then converted into actual voltages via a (1,1)-type cross-talk tensor $\tens{Q}^p_q$ (Eq.~(\ref{eq:gme-deriv-Qv})).

The underlying assumption here is that each receptor's actual voltage can be derived from the nominal voltages alone. To see why this may
be simplistic, consider a hypothetical array of of $n$ identical dipoles in the same orientation, parallel to the
$x$ axis. Nominally, the dipoles are then only sensitive to the $x$ component of the EMF, which, in terms of the PB tensor $\tens{E}$, means that
$\tens{E}_{2}^{p}\equiv0$ for all $p$. Consequently, the actual voltages $v^{p}$ given by this model will only depend on the $x$ component of the EMF. \EDIT{If mutual coupling causes any dipole to be sensitive to the $y$ component of the EMF seen at another dipole, this results in a contamination of the measured signal that cannot be described by this voltage-only cross-talk model.}

A more general approach is to describe the voltage response of each receptor as a function of the EMF at all the receptor locations, rather than the
nominal receptor voltages. This requires a (1,2)-type tensor:

\[
v^{p}=\tens{E}_{\sigma\beta}^{p}e'^{\sigma\beta}.
\label{eq:v-tensor-gen}
\]

This $\tens{E}^p_{ij}$ tensor (consisting of $n\times n \times 2$ complex numbers) then describes the PB response and the mutual 
coupling together. The simpler cross-talk-only model corresponds to $\tens{E}^p_{qj}$ being decomposable into a product of two
(1,1)-type tensors ($n\times n+ n\times2$ complex numbers), as $\tens{E}_{qj}^{p}=\tens{Q}_q^p \tens{E}_j^p.$ This model will perhaps prove to be sufficient in real-life applications, but it is illustrative how simple it is to extend the formalism to the more complex case.

\subsection{Describing beamforming}

In modern PAF and AA designs, receptors are grouped into \emph{stations}, and operated in beamformed mode -- that is, groups of receptor 
voltages are added up with complex weights to form one or more \emph{compound beams}. The output of a station is then a single complex voltage 
(strictly speaking, a single complex number, since beamforming is usually done after A/D conversion) per each compound beam, rather than $n$ individual 
receptor voltages.

Beamforming may also be described in terms of the tensor RIME. Let's assume $N$ stations, each being an array of $n_{1},n_{2},...,n_{N}$
receptors. The voltage vector $\vec v^a$ registered at station $p$ (where $a=1...n_p$) can be described by Eq.~(\ref{eq:v-tensor}). In addition, the voltages are subject to per-receptor complex gains (which we had quietly ignored up until now), which corresponds to another term, $g^a$.
The output of one beamformer, $f$, is computed by multiplying this by a covector of weights, $w_{a}$:

\begin{equation}
\label{eq:beamformer-output}  
f=w_{a}v^{a}=w_a g^a \tens{Q}_{\sigma}^{a}\tens{E}_{\beta}^{\sigma}\tens{K}^{\sigma}\tens{Z}_{\alpha}^{\sigma\beta}e^{\alpha}.
\end{equation}

In a typical application, the beamformer outputs are correlated across stations. In this context, it is useful to derive a compound beam tensor,
which would allow us to treat a whole station as a single receptor. To do this, we must assume that image plane effects are the same for
all receptors in a station ($\tens{Z}_{\alpha}^{\sigma\beta}\equiv \tens{Z}_{\alpha}^{\beta}$). Furthermore, we need to decompose the 
phase term $\tens{K}^{\sigma}$ into a common ``station phase'' $\tens{K}$, and a per-receptor differential delay $(\tens{\delta{K)}}^{\sigma}$, so that $\tens{K}^{\sigma}=\tens{K}{(\tens{\delta K})}^{\sigma}$. The latter can be derived in a straightforward way from the station (or dish) geometry. We can then collapse some summation indices:

\begin{eqnarray}
\nonumber f & = &w_{a}g^a\tens{Q}_{\sigma}^{a}\tens{E}_{\beta}^{\sigma}\tens{K}{\tens{(\delta K)}}^{\sigma}\tens{Z}_{\alpha}^{\beta}e^{\alpha}=
(w_{a}g^a\tens{Q}_{\sigma}^{a}\tens{E}_{\beta}^{\sigma}{\tens{(\delta K)}}^{\sigma})\tens{K}\tens{Z}_{\alpha}^{\beta}e^{\alpha} \\
& = & \tens{S}_{\beta}\tens{K}\tens{Z}_{\alpha}^{\beta}e^{\alpha}.
\label{eq:compound-beam}
\end{eqnarray}

This expression is quite similar to Eq.~(\ref{eq:v-tensor}). Now, if for station $p$ the compound 
beam tensor is given by $\tens{S}_{\beta}^{p}$, then a complete RIME for an interferometer composed of beamformed stations is:

\begin{equation}
\tens{V}_{q}^{p}=\tens{S}_{\beta}^{p}\tens{K}^{p}\tens{Z}_{\alpha}^{a\beta}\tens{B}_{\delta}^{\alpha}\bar\tens{Z}_{q\gamma}^{\delta}\bar\tens{K}_{q}\bar\tens{S}_{q}^{\gamma},\label{eq:gme-beamformer}
\end{equation}

which is very similar to the RIME of Eq.~(\ref{eq:gme0}), except that
the PB tensor $\tens{E}$ has been replaced by the station beam tensor
$\tens{S}$, and there's no cross-talk between stations. If each station produces a pair of compound beams (e.g.,
for the same pointing but sensitive to different polarizations), then this equation reduces to the classical matrix RIME, where the $E$-Jones term is given by a tensor product. In principle, we could also combine Eqs.~(\ref{eq:beamformer-output}) and (\ref{eq:gme-beamformer}) into one long equation describing both beamforming and station-to-station correlation.

This shows that a compound beam tensor ($\tens{S}_{i}^{p}$) can always be derived from the beamformer weights, receptor gains, mutual coupling terms, element PBs, and station geometry, under the assumption that image-plane effects are the same across the aperture of the station. By itself this fact is not particularly new or surprising, but its useful to see how the tensor formalism allows it to be formulated as an integral part of the RIME.

As for the image-plane effects assumption, it is probably safe for PAFs and small AAs, but perhaps not so for large AAs. If the assumption does not hold, we're left with an extra $\sigma$ index in Eq.~(\ref{eq:compound-beam}), and may no longer factor out an independent compound beam tensor 
$\tens{S}_{i}^{p}$. This situation cannot be described by the Jones formalism at all, but is easily accommodated by the tensor RIME.

\subsection{A classical dual-pol RIME}

Equation~(\ref{eq:gme0}) describes all correlations in an interferometer in a single (1,1)-type tensor (matrix). Contrast this to Eq.~(\ref{eq:me0-tensor1}), which does the same via a (2,2)-type tensor, by grouping pairs of receptors per station. Since the latter is a more familiar form in radio interferometry, it may be helpful to recast Eq.~(\ref{eq:gme0}) in the same manner. First, we mechanically replace each receptor index ($p,q,\sigma,\tau$) by pairs of indices ($pi$, $qj$, $\sigma\upsilon$, $\tau\phi$), corresponding to a station and a receptor within the station:

\[
\tens{V}^{pi}_{qj} = 
  \tens{Q}_{\sigma\upsilon}^{pi}\tens{E}_{\beta}^{\sigma\upsilon}\tens{K}^{\sigma\upsilon}\tens{Z}_{\alpha}^{\sigma\upsilon\beta}
  \tens{B}^\alpha_\delta
  \bar\tens{Z}_{\tau\phi\gamma}^{\delta}\bar\tens{K}_{\tau\phi}\bar\tens{E}_{\tau\phi}^{\gamma}\bar\tens{Q}_{qj}^{\tau\phi}.
\]

Next, we assume colocation (since the per-station receptors are, presumably, colocated) and simplify some tensors. In particular, 
$\tens{K}$ and $\tens{Z}$ can lose their receptor indices:

\[
\tens{V}^{pi}_{qj} = 
  \tens{Q}_{\sigma\upsilon}^{pi}\tens{E}_{\beta}^{\sigma\upsilon}\tens{K}^{\sigma}\tens{Z}_{\alpha}^{\sigma\beta}
  \tens{B}^\alpha_\delta
  \bar\tens{Z}_{\tau\gamma}^{\delta}\bar\tens{K}_{\tau}\bar\tens{E}_{\tau\phi}^{\gamma}\bar\tens{Q}_{qj}^{\tau\phi}.
\]

This equation cannot as yet be expressed via the Jones formalism, since for any $p,q$, the sum on the right-hand side contains terms for other stations 
($\sigma,\tau$). To get to a Jones-equivalent formalism, we need to remember the closed system assumption, i.e. no cross-talk or mutual coupling between stations (Sect.~\ref{sec:closed-sys-assumption}). This corresponds to $\tens{Q}^{pi}_{qj}\equiv0$ for $p\ne q$. $\tens{Q}$ is then equivalent to a tensor of one rank lower, with one station index eliminated:

\begin{equation}
\label{eq:gme-classic-xtalk}
\tens{V}^{pi}_{qj} = 
  \tens{Q}_{\upsilon}^{pi}\tens{E}_{\beta}^{p\upsilon}\tens{K}^{p}\tens{Z}_{\alpha}^{p\beta}
  \tens{B}^\alpha_\delta
  \bar\tens{Z}_{q\gamma}^{\delta}\bar\tens{K}_{q}\bar\tens{E}_{q\phi}^{\gamma}\bar\tens{Q}_{qj}^{\phi}.
\end{equation}

For any $p,q$, this is now exactly equivalent to a Jones-formalism RIME of the form:

\[
  \coh{V}{pq} = \jones{Q}{p} \jones{E}{p} K_p \jones{Z}{p} \coh{B}{} \jonesT{Z}{q} K^\herm_q \jonesT{E}{p} \jonesT{Q}{q},
\]
where $K_p$ is a scalar, and the rest are full $2\times2$ matrices. The $\jones{Q}{p}$ term here incorporates the traditional $G$-Jones (receiver gains) and $D$-Jones (polarization leakage). Finally, if we assume no polarization leakage (i.e. no cross-talk between receptors), then $\tens{Q}^{pi}_{j}\equiv0$ for $i\ne j$, and we can lose another index:

\begin{equation}
\label{eq:gme-classic-no-xtalk}
\tens{V}^{pi}_{qj} = 
  \tens{Q}^{pi}\tens{E}_{\beta}^{pi}\tens{K}^{p}\tens{Z}_{\alpha}^{p\beta}
  \tens{B}^\alpha_\delta
  \bar\tens{Z}_{q\gamma}^{\delta}\bar\tens{K}_{q}\bar\tens{E}_{qj}^{\gamma}\bar\tens{Q}_{qj}.
\end{equation}

In the Jones formalism, this is equivalent to $\jones{Q}{p}$ being a diagonal matrix for any given $p$.

\subsection{A full-sky tensor RIME}

By analogy with the matrix RIME \citep[see Paper I,][Sect.~3]{RRIME1}, we can extend the tensor formalism to the full-sky case. This does not lead to any new insights at present, but is given here for the sake of completeness.

When observing a real sky, each receptor is exposed to the superposition of the EMFs arriving from all possible directions. Let's begin with Eq.~(\ref{eq:gme0}), and assume the $\tens{Q}$ term is a DIE, and the rest are DDEs.  For a full-sky RIME, we need to integrate the equation over all directions as

\[
\tens{V}_{q}^{p}=\tens{Q}_{\sigma}^{p}\left(
\int\limits_{4\pi} \tens{E}_{\beta}^{\sigma}\tens{K}^{\sigma}\tens{Z}_{\alpha}^{\sigma\beta}\tens{B}_{\delta}^{\alpha}
\bar\tens{Z}_{\tau\gamma}^{\delta}\bar\tens{K}_{\tau}\bar\tens{E}_{\tau}^{\gamma}\DD{\Omega}
\right)\bar\tens{Q}_{q}^{\tau},
\]

which, projected into $lm$ coordinates, gives us:

\begin{equation}
\tens{V}_{q}^{p}=\tens{Q}_{\sigma}^{p}\left(
\iint\limits_{lm} \tens{E}_{\beta}^{\sigma}\tens{K}^{\sigma}\tens{Z}_{\alpha}^{\sigma\beta}\tens{B}_{\delta}^{\alpha}
\bar\tens{Z}_{\tau\gamma}^{\delta}\bar\tens{K}_{\tau}\bar\tens{E}_{\tau}^{\gamma}\frac{\DD{l}\DD{m}}{n}
\right)\bar\tens{Q}_{q}^{\tau},
\label{eq:gme-fullsky-int}
\end{equation}

Let's isolate a few tensor sub-products and collapse indices. First, we can introduce an ``apparent
sky'' tensor: 

\[
\hat\tens{B}^p_q = 
\tens{E}_{\beta}^p \tens{Z}_{\alpha}^{p\beta} \tens{B}_{\delta}^{\alpha}
\bar\tens{Z}_{q\gamma}^{\delta}\bar\tens{E}_{q}^{\gamma}
\]

Note that this is an $n\times n$ matrix. Physically, $\tens{B}^p_q(l,m)$ corresponds to the coherency ``seen'' by receptors $p$ and $q$ as a function of direction. Next, we introduce a phase
tensor:

\[
\tens{K}^p_q =\tens{K}^p\bar\tens{K}_q,
\]

which another $n\times n$ matrix. Note that we reuse the letter K here, but there shouldn't be any confusion with
with the ``other'' $\tens{K}^p$, since the tensor type is different. Each component of this tensor is given by

\[
\tens{K}^p_q = \exp \left  [ -2\pi i (u_{pq}l+v_{pq}m+w_{pq}(n-1)) \right ].
\]

Equation (\ref{eq:gme-fullsky-int}) then becomes simply:

\begin{equation}
V_{q}^{p}=Q_{\sigma}^{p}\left(\iint\limits_{lm}B_{\tau}^{\sigma}K_{\tau}^{\sigma}\frac{\DD{m}\DD{m}}{n}\right)\bar{Q}_{q}^{\tau},
\label{eq:gme-fullsky-int1}
\end{equation}

where the integral then corresponds to $n\times n$ element-by-element Fourier transforms, and all the DDE-related discussions of Papers~I \citep[Sect.~3]{RRIME1} and II \citep[Sect.~2]{RRIME2} apply.

\subsubsection{The apparent coherency tensor}

If we designate the value of the integral in Eq.~(\ref{eq:gme-fullsky-int}) by the \emph{apparent coherency tensor} $\tens{X}_{\tau}^{\sigma}$, we have
arrive at the simple equation

\[
\tens{V}_{q}^{p}=\tens{Q}_{\sigma}^{p}\tens{X}_{\tau}^{\sigma}\bar\tens{Q}_{q}^{\tau}.
\]

which ties together the observed correlation matrix, $\tens{V}_{q}^{p}$, and the apparent coherency tensor $\tens{X}_{\tau}^{\sigma}$. The physical
meaning of each element of $\tens{X}_{\tau}^{\sigma}$ is, obviously, the apparent coherency observed by receptor pair $\sigma$ and $\tau$.
The cross-talk term $\tens{Q}$ {}``scrambles up'' the apparent coherencies among all receptors. Note that this similar to 
the coherency matrix $\tens{X}$ (or $\coh{X}{pq}$) used in the classical formulation of the matrix RIME \citep[Sect.~1.7]{ME1,RRIME1}.

%
%
%
%
%

\subsection{Coordinate transforms, or whither tensors?}

Einstein summation by itself is a powerful notational convenience for expressing linear combinations of multidimensional arrays, one that can be gainfully employed without regard to the finer details of tensors. The formulations of this paper may in fact be read in just such a manner, especially as they do not seem to make explicit use of that many tensor-specific constructs. It is then a fair question whether we need to invoke the deeper concepts of tensor algebra at all.

There is one tensor property, however, that is crucial within the context of the RIME, and that is behaviour under coordinate transforms. In the formulations above, I did not specify a coordinate system. As in the case of the matrix RIME, the equations hold under change of coordinate frame, but their components must be transformed following certain rules. The general rules for tensors are given in Sect.~\ref{sec:tensors} (Eq.~(\ref{eq:tensor-coord-xform})); for the mixed-dimension tensors employed in this paper, coordinate transforms only affect the core $\mathbb{C}^2$ (or $\mathbb{C}^3$) vector space, and do not apply to station indices (the latter are said to be \emph{invariant}, see Sect.~\ref{sec:tensors-mixed-dim}).

As long as we know that something is a tensor of a certain type, we have a clear rule for coordinate transformations given by Eq.~(\ref{eq:tensor-coord-xform}). However, Einstein notation can be employed to form up arbitrary expressions, which are not necessarily proper tensors unless the rigorous rules of tensor algebra are followed (see Appendix~\ref{sec:tensor-theory}). This argues against a merely mechanical use of Einstein summation, and makes it worthwhile to maintain the mathematical rigour that enables us to clearly follow whether something is a tensor or not.

\section{Implementation aspects}

Superficially, evaluation of Einstein sums seems straightforward to implement in software, since it is just a series of nested loops.
Upon closer examination, it turns out to raise some non-trivial performance and optimization issues, which I'll look at in this section.

\subsection{A general formula for FLOP counts}

Consider an Einstein sum that is a product of $k$ tensors (over a $D$-dimensional vector space), with $n$ free and $m$ summation indices. I'll call this an ($n,m,k)$-\emph{calibre\/} product. Let's count the number of floating-point operations (\emph{ops\/} for short) required to compute the result. The resulting tensor has $D^n$ components. Each component is a sum of $D^m$ individual products (thus $D^m-1$ additions); each product incurs $k-1$ multiplications. The total op count is thus 

\begin{equation}
N_\mathrm{ops}^{(D)}(n,m,k) = D^n((D^m(k-1)+D^m-1)) = D^n(D^m k -1 ).
\end{equation}

For mixed-dimensionality tensors, a similar formula may be derived by replacing $D^n$ and $D^m$ with $D_1^{n_1}D_2^{n_2}$ and $D_1^{m_1}D_2^{m_2}$, where the two dimensions are $D_1$ and $D_2$, and the index counts per each dimensionality are numbered accordingly.

Consider a few familiar examples:

\begin{itemize}
\item Multiplication of $2\times2$ matrices, $\tens{A}^i_\alpha\tens{B}^\alpha_j$: $N_\mathrm{ops}^{(2)}(2,1,2)=12$.
\item Multiplication of $4\times4$ matrices: $N_\mathrm{ops}^{(4)}(2,1,2)=112$.
\item Outer product of $2\times2$ matrices, $\tens{A}^i_j\tens{B}^k_l$:  $N_\mathrm{ops}^{(2)}(4,0,2)=16$.
\item Multiplication of a $4\times4$ matrix by a 4-vector, $\tens{A}^i_\alpha x^\alpha$:  $N_\mathrm{ops}^{(4)}(1,1,2)=28$.
\item The equivalent operation (see Eq.~(\ref{eq:me0-tensor-Jpq})) of multiplying a (2,2)-type tensor (with $D=2$) by a matrix, 
$\tens{J}^{i\beta}_{j\alpha}\tens{B}^\alpha_\beta$: $N_\mathrm{ops}^{(2)}(2,2,2)=28$.

\end{itemize}

\subsection{Partitioning an Einstein sum}

Mathematically equivalent formulations can often incur significantly different numbers of ops. For example, in Paper~I \citep[Sect.~6.1]{RRIME1}, I already noted that a straightforward implementation of a $2\times2$ RIME is cheaper than the same equation in $4\times4$ form, although the specific op counts given therein are in error\footnote{Specifically, Paper~I claims 128 ops per Jones term in the $4\times4$ formalism: 112 to multiply two $4\times4$ matrices, and another 16 for the outer product. These numbers are correct per se. However, a $4\times4$ RIME may in fact be evaluated in a more economical order, namely as a series of multiplications of a 4-vector by a $4\times4$ matrix. As seen above, this costs 28 ops per each matrix-vector product, plus 16 for the outer product, for a total of only 44 ops per Jones term.}.

Let's consider a $2\times2$ RIME of the form of Eq.~(\ref{eq:rime-onion}), with two sets of Jones terms, which we'll designate as $\jones{D}{}$ and $\jones{E}{}$. We then have the following fully-equivalent formulations in $2\times2$ and $4\times4$ form:

\begin{eqnarray}
\label{eq:example-2x2}
  \coh{V}{pq} &=& \jones{D}{p}\jones{E}{p} \coh{B}{} \jonesT{E}{q} \jonesT{D}{q}, \\
\label{eq:example-4x4}
  \vec v_{pq} &=& (\jones{D}{p}\otimes\jonesT{D}{q})(\jones{E}{p}\otimes\jonesT{E}{q})
              \mathbf{S}{} \vec I.
\end{eqnarray}

while in tensor notation the same equation can be formulated as

\begin{equation}
  [\tens{V_{pq}}]^i_j = 
  [\tens{D_p}]^{i}_{\alpha_2} 
  [\tens{E_p}]^{\alpha_2}_{\alpha_1} 
  \tens{B}^{\alpha_1}_{\beta_1}
  [\bar\tens{E_q}]^{\beta_1}_{\beta_2}  
  [\bar\tens{D_q}]^{\beta_2}_{j}.
\label{eq:tensor-onion}
\end{equation}

The cost of this Einstein sum is $N_\mathrm{ops}^{(2)}(2,4,5)=316$ ops. In comparison, the $2\times2$ form incurs 4 matrix products, for a total of only 48 ops, while the $4\times4$ form incurs\footnote{Not counting the $\mathbf{S}\vec I$ operation: we assume the coherency vector is a given, since it's the equivalent of $\coh{B}{}$.} two outer products and two $4\times4$ matrix/vector products, for a total of 88. For longer expressions with more Jones terms (i.e. larger $m$), brute-force Einstein summation does progressively worse.

It is easy to see the source of this inefficiency. In the innermost loop (say, over $j$), only the rightmost term $[\bar\tens{D_q}]^{\beta_2}_j$ is changing, so it is wasteful to repeatedly take the product of the other four terms at each iteration. We can trim some of this waste by computing things in a slightly more elaborate order. Let's split up the computation as

\begin{equation}
  \underbrace{\left(
    [\tens{D_p}]^{i}_{\alpha_2} 
    [\tens{E_p}]^{\alpha_2}_{\alpha_1} 
    \tens{B}^{\alpha_1}_{\beta_1}
    [\bar\tens{E_q}]^{\beta_1}_{\beta_2}\right)
  }_{\equiv\tens{A}^i_{\beta_2}}
  [\bar\tens{D_q}]^{\beta_2}_{j},
\label{eq:tensor-partition}
\end{equation}

and compute $\tens{A}^i_{\beta_2}$ first (costing $N_\mathrm{ops}^{(2)}(2,3,4)=124$ ops), followed by $\tens{A}^i_{\beta_2}[\bar\tens{D_q}]_{j}^{\beta_2}$ (costing $N_\mathrm{ops}^{(2)}(2,1,2)=12$ ops). This is an improvement, but we don't have to stop here: a similar split can be applied to $\tens{A}^i_{\beta_2}$ in turn, and so on, ultimately yielding the following sequence of operations:

\begin{equation}
  \left(
  \left(
  \left(
    [\tens{D_p}]^{i}_{\alpha_3} 
    [\tens{E_p}]^{\alpha_3}_{\alpha_2}
  \right)
  \tens{B}^{\alpha_1}_{\beta_1}
  \right)
  [\bar\tens{E_q}]^{\beta_2}_{\beta_3}
  \right)
  [\bar\tens{D_q}]^{\beta_3}_{j}
\label{eq:tensor-partition-2x2})
\end{equation}

But this is just a sequence of $2\times2$ matrix products, i.e. exactly the same computations that occur in the $2\times2$ formulation! And on the other hand, as already intimated by Eq.~(\ref{eq:me0-tensor-Jpq}), the $4\times4$ form is equivalent to a different partitioning of the same expression, namely

\begin{equation}
  \left( 
    [\tens{D_p}]^i_{\alpha_3} [\bar\tens{D_q}]_{j}^{\beta_3} 
  \right) 
  \left(
    \left( 
      [\tens{E_p}]^{\alpha_3}_{\alpha_2} [\bar\tens{E_q}]_{\beta_3}^{\beta_2} 
    \right) 
      \tens{B}^{\alpha_1}_{\beta_1} 
  \right).
\label{eq:tensor-partition-4x4})
\end{equation}

The crucial insight here is that different partitionings of the computation in Eq.~(\ref{eq:tensor-onion}) incur different numbers of ops. 

Let's look at what happens to the calibres during partitioning. Consider a partition of an $(n,m,k)$-calibre product into two steps. The first step computes an $(n',m',k')$-calibre sub-product (for example, in Eq.~(\ref{eq:tensor-partition}), the initial calibre is $(2,4,5)$, and the sub-product for $\tens{A}^i_{\beta_2}$ has calibre $(2,3,4)$). At the second step, the result of this is multiplied with the remaining terms, resulting in an expression of calibre $(n,m-m',k-k'+1)$ (in Eq.~(\ref{eq:tensor-partition}), this is $\tens{A}^i_{\beta_2}\bar\tens{D}^{\beta_2}_j$, with a calibre of $(2,1,2)$). The calibres are strictly related: each of the $k$ terms goes to either one sub-product or the other, but we incur one extra term ($\tens{A}$) in the partitioning, hence we have $k-k'+1$ terms at the second step. The summation indices are also partitioned between the steps, hence $m-m'$ are left for the second step. As for the free indices, their number $n'$ may actually temporarily increase (as in the case of Eq.~(\ref{eq:tensor-partition-4x4}), where sub-products have the calibre (4,0,2)). It is straightforward to show that if it does not increase, then 

\[
  N_\mathrm{ops}^{(D)}(n,m',k') + N_\mathrm{ops}^{(D)}(n,m-m',k-k'+1) < N_\mathrm{ops}^{(D)}(n,m,k),
\]
so as long as $n'\leq n$, partitioning always reduces the total number of ops. In essence, this happens because in the total op counts, a product is replaced by a sum: $D^{m'}+D^{m-m'}<D^m$. 

From this it follows that the $2\times2$ form of the RIME is, in a sense, optimal. The partitioning given by Eq.~(\ref{eq:tensor-partition-2x2})
reduces an $N_\mathrm{ops}^{(D)}(2,m,k)$ operation into $m$ operations of $N_\mathrm{ops}^{(D)}(2,1,2)$ each; the latter represents the smallest possible non-trivial sub-product. (Note that the rank of any sub-product of Eq.~(\ref{eq:tensor-onion}) can only be an even number, since all the terms have an even rank. The minimum non-trivial $n'$ is therefore 2.)

\subsection{Dependence optimization}

The partitioning given by Eq.~(\ref{eq:tensor-partition-2x2}) allows for a few alternatives, corresponding to different order of matrix multiplication. While seemingly equivalent, they may in fact represent a huge opportunity for optimization, when we consider that in real life, 
the equation needs to be evaluated millions to billions of times, for all antenna pairs, baselines, time and frequency bins. Not all the terms in the equation have the same time and frequency dependence: some may be constant, some may be functions of time only or frequency only, some may change a lot slower than others -- in other words, some may have \emph{limited dependence}. For example, in the ``onion'' of Eq.~(\ref{eq:example-2x2}), if $\coh{B}{}$ and $\jones{E}{p}$ have a limited dependence, say on frequency only, then the inner part of the ``onion'' can be evaluated in a loop over frequency channels, but not over timeslots. The resulting savings in ops can be enormous.

This was already demonstrated in the MeqTrees system \citep{meqtrees}, which takes advantage of limited dependence on-the-fly. A RIME like Eq.~(\ref{eq:example-2x2}) is represented by a \emph{tree}, which typically corresponds to the following order of operations:

\[
  \coh{V}{pq} =  \jones{D}{p}(\jones{E}{p} \coh{B}{} \jonesT{E}{q} ) \jonesT{D}{q}, \\
\]
with an outermost loop being over the layers of the ``onion'', and inner loops over times and frequencies. When the operands have limited dependence (as e.g. for the $\jones{E}{p} \coh{B}{} \jonesT{E}{q}$ product), the MeqTrees computational engine automatically skips unnecessary inner loops. Thus the amount of loops is minimized on the ``inside'' of the equation, and grows as we move ``out'' through the layers and add terms with more dependence. I call this \emph{dependence optimization}. 

Among the alternative partitionings of Eq.~(\ref{eq:tensor-partition-2x2}), the one that computes the sub-products with the least dependence first can benefit from dependence optimization the most.

\subsection{Commutation optimization}

In a $2\times2$ matrix RIME, dependence optimization works best if the terms with the least dependence are placed on the inside of the equation -- if limited dependence happens to apply to $\jones{D}{p}$ (on the outside) and not $\jones{E}{p}$ (on the inside), dependence optimization can't reduce any inner loops at all. Unfortunately, one cannot simply change the order of the matrices willy-nilly, since they don't generally commute. However, when dealing with specific kinds of matrices that \emph{do\/} commute, we \emph{can} do some optimization by shuffling them around. Real-life RIMEs tend to be full of matrices with limited commutation properties, such as scalar, diagonal and rotation matrices \cite[see Paper~I,][Sect.~1.6]{RRIME1}.

In tensor notation, $\tens{A}^i_\alpha \tens{B}^\alpha_j$ commute if the summation indices can be swapped around: $\tens{A}^i_\alpha \tens{B}^\alpha_j=\tens{A}^\alpha_j \tens{B}^i_\alpha.$  Some of the commutation rules of $2\times2$ matrices discussed in Paper~I (\emph{ibid.}) do map to tensor form easily. For example, a diagonal matrix is a tensor with the property $\tens{A}^i_j=0$ for $i\ne j$. If $\tens{A}$ and $\tens{B}$ are both diagonal, then $\tens{A}^i_\alpha \tens{B}^\alpha_j$ is only non-zero for $i=j$, and commutation is obvious. Other commutation rules are more opaque: rotation matrices are known to commute among themselves, but this does not follow from tensor notation at all.
Therefore, opportunities for commutation optimization of a tensor expression may not be detectable until we recast it into matrix form. 

\subsection{Towards automatic optimization}

For the more complicated expressions of e.g. Eq.~(\ref{eq:gme0}) or (\ref{eq:gme-beamformer}), different partitionings may prove to be optimal. The previous sections show how to do such analysis formally. In fact, one could conceive of an algorithm that, given a a tensor expression in Einstein form, searches for an optimal partitioning automatically (with dependences taken into account).

It is also possible that alternative partitionings may prove to be more or less amenable to parallelization and/or implementation on GPUs. The tensor formalism may prove to be a valuable tool in this area. Recasting a series of tensor operations (matrix products, etc.) as a single Einstein sum, such as that in Eq.~(\ref{eq:tensor-onion}), shows the computation in its ``flattest'' (if relatively inefficient) form, which can then be repartitioned to yield equivalent but more efficient formulations.

\section{Conclusions}

The $2\times2$ matrix RIME, having proven itself as a very capable tool for describing classical radio interferometry, is showing its limitations when confronted with PAFs, AAs, and wide-field polarimetry. This is due to a number of implicit assumptions underlying the formalism (plane-polarized, dual, colocated receptors, closed systems) that have been explicated in this paper. The RIME may be rewritten in an array correlation matrix form that makes these assumptions clearer, but this struggles to combine image-plane effects and mutual coupling in the same equation.

A more general formalism based on tensors and Einstein notation is proposed. This reduces to the $2\times2$ (and $4\times4$) forms of the RIME under the explicated assumptions. The tensor formalism can be used to step outside the bounds of these assumptions, and can accommodate regimes not readily described in $2\times2$ matrix form. Some examples of the latter are:

\begin{description}

\item[{\bf Coupling}] between closely-packed stations cannot be described in the basic $2\times2$ form (where a Jones chain corresponding to each signal path is used) without additional extrinsic complexity, such as combining the $2\times2$ equations into some kind of larger equation. The tensor formalism describes this regime with a single equation.

\item[{\bf Beamforming}] can only be accommodated in the $2\times2$ form by using separate equations to derive the effective Jones matrix of a beamformer. The tensor formalism combines beamforming and correlation into the same equation.

\item[{\bf The Wolf formalism}] proposed by \citet{Carozzi:ME3D} uses $3\times3$ matrices to describe polarimetry in the wide-field regime. Again, this can only be mapped to the $2\times2$ form by using external equations to derive special Jones matrices. A tensor formalism naturally incorporates the 3-component description, and can be used to combine it with the regimes above.

\end{description}

In practice, tensor equations may be implemented via alternative formulations that are mathematically equivalent, but have significantly different 
computing costs (the $2\times2$ vs. $4\times4$ RIMEs being but one example). Computing costs may be optimized by a repartitioning of the calculations, and some formal methods for analysing this have been proposed in this paper.

I do not propose to completely supplant the $2\times2$ RIME. Where applicable (that is, for the majority of current radio interferometric observations), the latter ought to remain the formalism of choice, both for its conceptual simplicity, and its computational efficiency. Even in these cases, the tensor formalism can be of value as both a rigorous theoretical tool for analysing the limits of the Jones formalism's applicability, and a practical tool for deriving certain specialized Jones matrices. 

\begin{acknowledgements}

This effort/activity is supported by the European Community Framework Programme 7, Advanced Radio Astronomy in Europe, grant agreement no.: 227290.

\end{acknowledgements}

\appendix

\section{Elements of tensor algebra}
\label{sec:tensor-theory}

This appendix gives a primer on elements of tensor theory relevant to the present paper. Most of this is just a summary of existing mathematical theory; the only new results are presented in Sect.~\ref{sec:mapping-rime-tensors}, where I formally map the RIME into tensor algebra. More details on tensor theory can be found in any number of textbooks, for example \citet{synge-schild-tensor} and \citet{simmonds-tensor-analysis}. 

As a preliminary remark, one should keep in mind that there are, broadly, two complementary ways of thinking about linear and tensor algebra. The \emph{coordinate} approach defines vectors, matrices, tensors, etc., in terms of their coordinate components in a particular coordinate system, and postulates somewhat mechanistic rules for transforming these components under change of coordinate frames. The \emph{intrinsic} (or abstract) approach defines these objects as abstract entities (namely, various kinds of linear functions) that exist independently of coordinate systems, and derives rules for coordinate manipulation as a result. For example, in the coordinate approach, a matrix is defined as an $n\times m$ array of numbers. In the intrinsic approach, it is a linear function mapping rank-$m$ vectors to rank-$n$ vectors.

Historically, the coordinate approach was developed first, and favours applications of the theory (which is why it is prevalent in e.g. physics and engineering). The intrinsic approach is favoured by theoretical mathematicians, since it has proven to be a more powerful way of extending the theory. When applying the theory to a new field (e.g. to the RIME), the mechanistic rules may be a necessity (especially when it comes to software implementations), but it is critically important to maintain conceptual links to the intrinsic approach, since that is the only way to verify that the application remains mathematically sound. This appendix therefore tries to explain things in terms of both approaches.

\subsection{Einstein notation}
\label{sec:einstein-notation}

Einstein notation uses upper and lower indices to denote components of multidimensional entities. For example, $x^i$ refers to the $i$-th component  of the vector $\vec x$ (rather than to $x$ to the power of $i$!), and $y_i$ refers to the $i$-th component of the \emph{covector} (see below) $\vec y$. Under the closely-related \emph{abstract index notation}, $x^i$ may be used to refer to $\vec x$ itself, with the index $i$ only serving to indicate that $\vec x$ is a one-dimensional object. Whether $x^i$ refers to the whole vector or to its $i$-th component is usually obvious from context (i.e. from whether a specific value for $i$ has been implied or not).

In general, upper indices are associated with \emph{contravariant} components (i.e., those that transform contravariantly), while lower indices refer  to \emph{covariant} ones. The following section will define these concepts in detail.

{\em Einstein summation} is a convention whereby the same index appearing in both an upper and lower position implies summation. That is,

\[
  x^i y_i = \sum_{i=1}^{N} x^i y_i, \;\;\;\;\;\; A^i_i = \sum_{i=1}^{N} A^i_i.
\]

An index that is repeated in the same position (e.g. $x_iy_i$) does \emph{not} imply summation\footnote{This point is occasionally ignored in the literature, i.e. one will see $x_i y_i$ implying summation over $i$ as well. This is mathematically sloppy from a purist's point of view.}. I shall be using the summation convention from this point on, unless otherwise stated.

\subsection{Vectors, covectors, co- and contravariancy}
\label{sec:covectors}

A vector can be thought of (in the coordinate approach) as a list of $N$ scalars drawn from a scalar field $\mathbb{F}$ (e.g., the field of complex numbers, $\mathbb{C}$). The set of all such vectors forms up the vector space $\mathbb{V}=\mathbb{F}^N$. For example, the Jones formalism is based on the vector space $\mathbb{C}^2$, while the Wolf formalism is based on $\mathbb{C}^3$. 

The intrinsic definition of a vector is straightforward but somewhat laborious, and can be looked up in any linear algebra text, so I will not reproduce it here. Intuitively, this is just the familiar concept of a length and direction in $N$-dimensional space. It is important to distinguish the vector as an abstract geometrical entity, existing independently of coordinate systems, from the list of scalars making up its representation in a specific coordinate frame. The terminology is also unhelpfully ambiguous; I will try to use \emph{vector} by itself for the abstract entity, and \emph{column vector}, \emph{row vector} or \emph{components} when referring to its coordinates. 

A {\em coordinate frame} in vector space $\mathbb{V}$ is defined by a set of linearly independent basis vectors $\vec e_1,...,\vec e_N$. Given a coordinate frame, any  vector $\vec x$ can be expressed as a linear combination of the basis vectors: $\vec x = x^i \vec e_i$ (using Einstein notation). Each \emph{component} of the vector, $x^i$, is a scalar drawn from $\mathbb{F}$. In matrix notation, the representation of $\vec x$ is commonly written as a \emph{column vector:}

\[
  \vec x = \left ( \begin{array}{c} x^1\\ \vdots \\ x^N \end{array} \right )
\]

A \emph{covector} (or a {\em dual vector\/}) $\vec f$ represents a {\em linear function} from $\mathbb{V}$ to $\mathbb{F}$, i.e. a linear function $f(\vec x)$ that maps vectors to scalars (in other words, covectors {\em operate\/} on vectors). This can also be written as $f:\mathbb{V}\mapsto\mathbb{F}$. The set of all covectors forms the {\em dual space\/} of $\mathbb{V}$, commonly designated as $\mathbb{V}^*$. In a specific coordinate frame, any covector may be represented by a set of $N$ scalar components $f_i$, so that the operation  $f(\vec x)$ becomes a simple sum: $f(\vec x) = x^if_i.$ In matrix notation, covectors are written as \emph{row vectors\/}:

\[
\vec f = \left( f_1 \, ... \, f_N \right), 
\]

and the operation $f(\vec x)$ is then just the matrix product of a row vector and a column vector, $\vec f\vec x$. Note that this definition is completely symmetric: we could as well have postulated a covector space, and defined the vector as a linear function mapping covectors onto scalars.

\subsubsection{Coordinate transforms}
\label{sec:coordinatre-xforms}

In the coordinate approach, both vectors and covectors are represented by $N$ scalars: the crucial difference is in how these numbers change under coordinate transforms. Consider two sets of basis vectors, $E=\{\vec e_i\}$ and $E'=\{\vec e'_i\}$. Each vector of the basis $E'$ can be represented by a linear combination of the basis $E$, as $\vec e'_j = a^i_j \vec e_i$. The components $a^i_j$ form the {\em transformation matrix} $\mathbf{A}$, which is an $N\times N$, invertible matrix. The change of basis can also be written as a matrix-vector product, by laying out the symbols for the basis vectors in a column, and formally following the rules of matrix multiplication:

\[
    \left( \begin{array}{c} \vec e'_1 \\ \vdots \\ \vec e'_N \end{array} \right ) = 
    \left( \begin{array}{ccc} 
      a^1_1 & \cdots & a^1_N \\
      \vdots & & \vdots \\
      a^N_1 & \cdots & a^N_N 
    \end{array} \right )  
    \left( \begin{array}{c} \vec e_1 \\ \vdots \\ \vec e_N \end{array}  \right ).
\]

Let $\vec x$ and $\vec x'$ be two column vectors representing the same vector in basis $E$ and $E'$, respectively; let $\vec f$ and $\vec f'$ be two row vectors representing a covector.

The crucial bit is this: in order for $\vec x$ and $\vec x'$ to represent the same vector in both coordinate systems, their components must transform \emph{contravariantly}, that is, as

\[
  \vec x' = \mathbf{A}^{-1} \vec x,\;\;\;\mbox{or}\;\;\;x'^i = \tilde a^i_j x^j,
\]
(in matrix or Einstein notation, respectively), where $\tilde a^i_j$ are the components of $\mathbf{A}^{-1}.$ In other words, the components of a vector transform in an opposite way to the basis\footnote{For a simple example, consider a basis $E'$ that is simply $E$ scaled up by a factor of 2: $e'_i = 2e_i$. In the $E'$ frame, a vector's coordinates will be {\em half} of those in $E$.}, hence \emph{contra\/}variantly.

On the other hand, in order for $\vec f$ and $\vec f'$ to represent the same covector (i.e. the same linear functional), their components must transform \emph{covariantly:}
\[
  \vec f' = \vec f \mathbf{A},\;\;\;\mbox{or}\;\;\;f'_i = a^i_j f_j.
\]

Vectors and linear functions (or covectors) are the two fundamental ingredients of the RIME.

\subsection{Vector products and metrics}

\subsubsection{Inner product}
\label{sec:inner-prod}

\newcommand{\inprod}[2]{\langle #1,#2 \rangle}

An \emph{inner product} on the vector space $\mathbb{R}^N$ or $\mathbb{C}^N$ is a function that maps two vectors onto a scalar. It is commonly designated as $\inprod{\vec x}{\vec y}=c$. Any function that is (1) linear, (2) conjugate-symmetric ($\inprod{\vec x}{\vec y}=\inprod{\vec y}{\vec x}^*$; for vector spaces over $\mathbb{R}$ this is simply symmetric), and (3) positive-definite ($\inprod{\vec x}{\vec x}>0$ for all $x\ne0$) can be adopted as an inner product. 

The dot product on Euclidean space $\mathbb{R}^N$ \[ \vec x\cdot \vec y = \sum_{i=1}^{N}x^iy^i, \] is an example of an inner product. In fact, since this paper (and other RIME-related literature) already uses angle brackets to denote averaging in time and frequency, I shall instead use the dot notation for inner products in general.

In matrix notation, the general form of an inner product on $\mathbb{C}^N$ spaces is $\vec x\cdot \vec y = \vec y^\herm \mathbf{M} \vec x$, where $\mathbf{M}$ is a positive-definite Hermitian $N\times N$ matrix. In order for this product to remain invariant under coordinate transform, the $\mathbf{M}$ matrix must transform as $\mathbf{M}'=\mathbf{A}^\herm \mathbf{M} \mathbf{A}$ (where $\mathbf{A}$ is the transformation matrix defined in the previous section), i.e. doubly-covariantly. Looking ahead to Sect.~\ref{sec:tensors}, this is an example of a (0,2)-type tensor. Another way to look at it is that the inner product defines a \emph{metric} on the vector space, and $\mathbf{M}$ gives the \emph{covariant metric tensor}, commonly designated as $\mathbf{M}=g_{ij}$. In Einstein notation, the inner product of $x^i$ and $y^j$ is then $g_{ij} x^i \overline{y^j}$, where $\overline{y^j}$ is the complex conjugate of $y^j$.

Given a coordinate system, any choice of Hermitian positive-definite $\mathbf{M}$ induces an inner product and a metric. In particular, choosing 
$\mathbf{M}$ to be unity results in a \emph{natural metric} of $\vec x \cdot \vec y=\vec y^\herm\vec x,$ which is in fact the one implicitly used in all RIME literature to date. Note, however, that if we change coordinate systems, the metric only remains natural if $\mathbf{A}$ is a unitary transformation ($\mathbf{A}^\herm \mathbf{A}=\mathbf{1}$). For example, a rigid rotation of the coordinate frame is unitary and thus preserves the metric, while a skew of the coordinates changes the metric. The other coordinate transform commonly encountered in the RIME, that between linear $xy$ and circularly-polarized $rl$ coordinates, is also unitary.

In tensor notation, the \emph{Kronecker delta} $\delta_{ij}$ ($\delta_{ij}=1$ for $i=j$, and 0 fo $i\ne j$), is often used to indicate a unity $\mathbf{M}$, i.e. as $\mathbf{M}=g_{ij}=\delta_{ij}.$

\subsubsection{Index lowering and conjugate covectors}
\label{sec:index-lowering}

An inner product induces a natural mapping (isomorphism) between $\mathbb{V}$ and $\mathbb{V}^*$, i.e. a pairing up of vectors and covectors.
For any vector $\vec y$, its \emph{conjugate covector} $\bar{\vec y}$ can be defined as the linear functional $\bar y(\vec w) = \vec w \cdot \vec y.$ On $\mathbb{C}^N$ space, this function is given by $\bar y(\vec w) = \vec y^\herm \mathbf{M} \vec w$, meaning simply that

\[
\bar{\vec y} = \vec y^H \mathbf{M},\;\;\;\mbox{or}\;\;\;\bar y_i=g_{ij} \overline{y^j},
\]

in matrix or Einstein notation, respectively. This operation is also known as \emph{index lowering}.
In a coordinate frame with the natural metric $\delta_{ij}$, index lowering is just the Hermitian transpose: $\bar{\vec y}\equiv\vec y^H$, or $\bar y_i=\overline{y^i}.$

For conciseness, this paper uses the notation $[y^i]^*$ or $\bar y_i$ (i.e. bar over symbol only) to denote the conjugate covector (or its $i$-th component) of the vector $\vec y$. Note how this is distinct from the complex conjugate of the $i$-th component, which is denoted by a bar over both the symbol and all indices, e.g. $\overline{y^i}.$

\subsubsection{Outer product}
\label{sec:outer-prod}

Given two vector spaces $\mathbb{V}$ and $\mathbb{W}$ and the dual space $\mathbb{W}^*,$ the \emph{outer product} of the vector 
$\vec x \in \mathbb{V}$ and the covector $\vec y^* \in \mathbb{W}^*$, denoted as $\mathbf{B}=\vec x\otimes\vec y^*$,  produces a linear transform between $\mathbb{W}$ and $\mathbb{V}$ (which, in other words, is a matrix), that is defined as:
\[
\mathbf{B}(\vec w) = \vec x y^*(\vec w),\;\;\;\mbox{or}\;\;\;[\mathbf{B}(\vec w)]^i = x^iy_jw^j
\]
i.e. the function given by $\vec y^*$ is applied to $\vec w$ (producing a scalar), which is then multiplied by the vector $\vec x$. 

Intrinsically, the outer product is defined on \emph{a vector and a covector}. If we also have an inner product on $\mathbb{W}$, we can use it to define the outer product operation on \emph{two vectors}, as $\vec x\otimes \vec y$ = $\vec x\otimes \bar{\vec y}$, where $\bar{\vec y}$ is the conjugate covector of $\vec y$ (see above).

Given a coordinate system in a complex vector space, the $\bar{\vec y}$ covector corresponds to the linear function $\bar{y}(\vec w) = \vec y^\herm \mathbf{M} \vec w,$ and the outer product $\mathbf{B}=\vec x\otimes \vec y$ is then
\[
\mathbf{B}(\vec w) = \vec x \vec y^\herm \mathbf{M} \vec w, \;\;\;\mbox{or}\;\;\;[\mathbf{B}(\vec w)]^i = g_{jk} x^i \overline{y^k} w^j.
\]
In other words, the outer product of $\vec x$ and $\vec y$ is given by the matrix product $\vec x \vec y^\herm \mathbf{M}$; in Einstein notation the corresponding matrix components are $b^i_j = g_{jk} x^i \overline{y^k}.$ 

Consider now a change of coordinates given by the transformation matrix $\mathbf{A}$. In the new coordinate system, the outer product becomes

$\mathbf{A}^{-1}\vec x \vec y^\herm \mathbf{M} \mathbf{A}, \;\;\;\mbox{or}\;\;\;b'^{l}_{m} = \tilde a_i^{l}a_m^{j} g_{jk} x^i \overline{y^k}$.

i.e. is transformed both contra- and covariantly. This is an example of a (1,1)-type tensor.

\subsection{Tensors}
\label{sec:tensors}

A tensor is a natural generalization of the vector and matrix concepts. In the coordinate approach, an \emph{$(n,m)$-type} tensor over the vector space $\mathbb{V}=\mathbb{F}^N$ is given by an $(n+m)$-dimensional array of scalars (from $\mathbb{F}$). A tensor is written using $n$ upper and $m$ lower indices, e.g.: $\tens{T}_{j_{1}j_{2}...j_{m}}^{i_{1}i_{2}...i_{n}}$. The \emph{rank} of the tensor is $n+m$. A vector (column vector) $x^i$ is a (1,0)-type tensor, a covector (row vector) $y_j$ is a (0,1)-type tensor, and a matrix is typically a (1,1)-type tensor. Note that the \emph{range} of each tensor index is, implicitly, from 1 to $N$, where $N$ is the rank of the original vector space. 

The upper indices correspond to contravariant components, and the lower indices to covariant components. Under a change of coordinates given by $\mathbf{A}=a^i_j$ ($\mathbf{A}^{-1}=\tilde a^i_j$), the components of the tensor transform $n$ times contravariantly and $m$ times covariantly: 

\begin{equation}
  \label{eq:tensor-coord-xform}
\tens{T'}_{j_1j_2...j_m}^{i_1i_2...i_n} = 
\tilde a^{i_1}_{k_1}\cdot...\cdot\tilde a^{i_n}_{k_n} \cdot 
a_{j_1}^{l_1}\cdot...\cdot a_{j_m}^{l_m} \cdot
\tens{T}_{l_1l_2...l_m}^{k_1k_2...k_n} 
\end{equation}

In the case of a Jones matrix (a (1,1)-type tensor), this rule corresponds to the familiar\footnote{Note that in Paper~I \citep[Sect.~6.3]{RRIME1} this shows up as $\mathbf{J}_T=\mathbf{T}\mathbf{J}_S\mathbf{T}^{-1}$, since the $\mathbf{T}$ matrix defined therein is exactly the inverse of $\mathbf{A}$ here.} matrix transformation rule of $\mathbf{J}'=\mathbf{A}^{-1}\mathbf{J}\mathbf{A}.$ Note that on the other hand, the metric $\mathbf{M}$ used to specify the inner product (Sect.~\ref{sec:inner-prod}) transforms differently, being a (0,2)-type tensor.

As far as typographical conventions go, this paper uses sans-serif capitals ($\tens{T}_j^i$) to indicate tensors in general, and lower-case italics ($x^i, y_j$) for vectors and covectors.

In the intrinsic (abstract) definition, a tensor is simply a linear function mapping $m$ vectors and $n$ covectors onto a scalar. This can be written as

\[
  \tens{T}: \underbrace{\mathbb{V}\times\cdots\times\mathbb{V}}_{m\;\mathrm{times}}\times
  \underbrace{\mathbb{V}^*\times\cdots\times\mathbb{V}^*}_{n\;\mathrm{times}} \mapsto \mathbb{F},
\]
or as $\tens{T}(\vec x_1,...,\vec x_m,\vec y^*_1,...,\vec y^*_n)=c.$ All the coordinate transform properties then follow from this one basic definition.

As a useful exercise, consider that a (1,1)-type tensor, which by this definition is a linear function $\tens{T}:\mathbb{V}\times\mathbb{V}^* \mapsto \mathbb{F}$, can be easily recast into as a linear transform of vectors. For any vector $\vec v$, consider the corresponding function $\vec v'(\vec w^*)$, operating on covectors, defined as $\vec v'(\vec w^*)= \tens{T}(\vec v,\vec w^*)$. This is a linear function mapping covectors to scalars, which as we know (see Sect~\ref{sec:covectors}) is equivalent to a vector. We have therefore specified a linear transform between $\vec v$ and $\vec v'$. On the other hand, we know that the latter can also be specified as a matrix -- and therefore a matrix is a (1,1)-type tensor. 

This line of reasoning becomes almost tautological if one writes out the coordinate components in Einstein notation. The result of $\tens{T}$ applied to the vector $v^i$ and the covector $w_i$ is a scalar given by the sum

\[
  \tens{T}(v^i,w_j) = \tens{T}^i_j v^j w_i = ( \tens{T}^i_j v^j ) w_i.
\]
Dropping $w_i$ and evaluating just the sum $\tens{T}^i_j v^j$ for each $i$ results in $N$ scalars, which are precisely the components of $v'^i$:
\[
  v'^i = \tens{T}^i_j v^j,
\]
which on the other hand is just multiplication of a matrix by a column vector, written in Einstein notation.

\subsubsection{Transposition and tensor conjugation}
\label{sec:tensor-transposition}

As a purely formal operation, transposition (in tensor notation) can be defined as a swapping of upper and lower indices:

\[
  [ v^i ]^T = v_i,\;\;\;[\tens{A}_{j}^{i}]^T = \tens{A}_{i}^{j},
\]
and the Hermitian transpose can be defined by combining this with complex conjugation. However, in the presence of a non-natural metric (Sect.~\ref{sec:inner-prod}), this is a purely mechanical operation with no underlying mathematical meaning, since it turns e.g. a vector into an entirely unrelated covector.

A far more meaningful operation is given by the index lowering procedure of Sect.~\ref{sec:index-lowering}, used to obtain conjugate covectors: $\bar v_i = g_{ij} \overline{v^j}$, and by its counterpart, {\em index raising\/}: $\bar w_i = g^{ij}\overline{w_j}$, where $g^{ij}$ is the \emph{contravariant metric tensor} (essentially the inverse of $g_{ij}$). This kind of \emph{tensor conjugation\/} can be generalized to the matrix case as:

\[
\bar\tens{A}^i_j = g^{i\alpha}g_{j\beta}\overline{\tens{A}_\alpha^\beta}.
\]
In the case of a natural metric $g_{ij} = \delta_{ij}$ (and only in this case), tensor conjugation is the same as a mechanical Hermitian transpose.

For conciseness, this paper uses the notation $\bar x_i$ to denote tensor conjugation, i.e. $\bar x_i$ is the conjugate covector of the vector $x^i$.

\subsection{Tensor operations and the Einstein notation}
\label{sec:tensor-ops}

Einstein notation allows for some wonderfully compact representations of linear operations on tensors, which result in other tensors. Some of these were already illustrated above:

{\bf Inner product: } $g_{ij} x^i \overline{y^j} = c$, resulting in a scalar.

{\bf Index lowering:} $\bar y_i = g_{ij} \overline{y^j}$, converting a (1,0)-type tensor (a vector) into a (0,1)-type tensor (its conjugate covector).

{\bf Outer product} of a vector and a covector, $b^i_j = x^i y_j$, resulting in a (1,1)-type tensor (a matrix). 

{\bf Matrix multiplication} of a matrix by a vector, resulting in another vector: $v'^i = \tens{T}^i_j v^j$.

Consider now multiplication of two matrices, which in Einstein notation can be written as
\[
\tens{A}_{j}^{i}=\tens{B}_{k}^{i}\tens{C}_{j}^{k}.\]
Here, $k$ is a \emph{summation} index (since it
is repeated), while $i$ and $j$ are \emph{free} indices. Free indices
propagate to the left-hand side of the expression. This is a very
easy formal rule for keeping track of what the type of the result
of a tensor operation is. For example, the result of $\tens{A}_{kl}^{ij}\tens{B}^{k}\tens{C}_{ij}$ is a (0,1)-type 
tensor, $d_{l}$ (i.e. just a humble covector), since $l$ is the only free index in the expression.

In complicated expressions, a useful convention is to use Greek letters for the summation indices, and Latin ones for the free indices:  $\tens{A}_{\theta}^{i}\tens{B}_{j}^{\theta}.$ This makes the expressions easier to read, but is not always easy to follow consistently. 
This paper tries to follow this convention as much as possible when recasting the RIME in tensor form.

The true power of Einstein notation is that it establishes relatively simple formal rules for manipulating tensor expressions. These rules
can help reduce complex expressions to manageable forms. 

\subsubsection{No duplicate indices in the same position}

Consider the expression $x^i y^i$. The index $i$ is nominally free, so can we treat the result as a (1,0)-type tensor $z^i = x^i y^i$? The answer is no, because $z^i$ does not transform as a (1,0)-type tensor. Under change of coordinates, we have

\[
  z'^i = x'^i y'^i = (\tilde a^i_j x^j) (\tilde a^i_k x^k) = \tilde a^i_j \tilde a^i_k x^j x^k, 
\]

which is not a single contravariant transform. In fact, $z^i$ transforms as a (2,0)-type tensor. 

Summation indices cannot appear multiple times either: the expression $\tens{W}^i_j = \tens{X}^i_\alpha \tens{Y}^\alpha \tens{Z}^\alpha_j$ is not a valid (1,1)-type tensor!

In general, any expression in Einstein notation will not yield a valid tensor if it contains repeated indices in the same (upper or lower) position. However, in this paper I make use of mixed-dimension tensors (Sect.~\ref{sec:tensors-mixed-dim}), with a restricted set of coordinate transforms, which results in some indices being effectively \emph{invariant}. Invariant indices can be repeated.

\subsubsection{Commutation}
\label{sec:einstein-commutation}
 
The terms in an Einstein sum \textbf{commute!} For any particular set of index values, each term in the sum represents a scalar, and the scalars as a whole make up one product in some complicated nested sum -- and scalars commute. For example, the matrix product $\tens{AB}=\tens{A}_\alpha^i \tens{B}_j^\alpha$ can be rewritten as $\tens{B}_{j}^{\alpha}\tens{A}_{\alpha}^{i}$ without changing the result. Were we to swap the matrices themselves around, the Einstein sum would become $\tens{B}_{\alpha}^{i}\tens{A}_{j}^{\alpha}\neq\tens{B}_{j}^{\alpha}\tens{A}_{\alpha}^{i}.$ Note how the relative position (upper vs. lower) of the summation index changes. In one case we're summing over columns of $\tens{B}$ and rows of $\tens{A}$, in the other case over columns of $\tens{A}$ and rows of $\tens{B}$. 

\subsubsection{Conjugation}

The tensor conjugate of a product is a product of the conjugates, with upper and lower indices swapped:

\begin{eqnarray*}
y^i = \tens{A}^i_\alpha x^\alpha, &\;\;\;& \bar y_i = \left[ \tens{A}^i_\alpha x^\alpha \right]^* = \bar\tens{A}_i^\alpha \bar x_\alpha. \\
\tens{C}^i_j = \tens{A}^i_\alpha \tens{B}^\alpha_j, &\;\;\;& \bar\tens{C}_i^j = \left[ \tens{A}^i_\alpha \tens{B}^\alpha_j \right ]^*= \bar\tens{A}_i^\alpha \bar\tens{B}_\alpha^j.
\end{eqnarray*}

This follows from the definition of conjugation in Sect.~\ref{sec:tensor-transposition}, and the commutation considerations above.

\subsubsection{Isolating sub-products and collapsing indices}

Summation indices can be ``collapsed'' by isolating intermediate products that contains all occurrences of that index.
For example, in the sum $\tens{A}_{\alpha}^{i}\tens{B}_{\beta}^{\alpha}\tens{C}_{j}^{\beta}$, the index $\beta$ can be collapsed by defining the intermediate product $\tens{E}_{j}^{\alpha}=\tens{B}_{\beta}^{\alpha}\tens{C}_{j}^{\beta}$. The sum then becomes simply a matrix product, $\tens{A}_{\alpha}^{i}\tens{E}_{j}^{\alpha}.$ It is important that the isolated sub-product contain \emph{all} occurrences of the index. For example, it would be formally incorrect to isolate the sub-product $\tens{F}_{j}^{\alpha}=\tens{B}_{\beta}^{\alpha}\tens{C}_{j}^{\beta}$, since the sum then become $\tens{A}_{\alpha}^{i}\tens{F}_{j}^{\alpha}\tens{D}^{\beta}$. The ``loose'' $\beta$ index on $\tens{D}$ then changes the type of the result.

\subsection{Mapping the RIME onto tensors}
\label{sec:mapping-rime-tensors}

This section contains some in-depth mathematical details (that have been glossed over in the main paper) pertaining to how the concepts of the RIME map onto formal definitions of tensor theory.

\subsubsection{Coherency as an outer product}
\label{sec:coh-outer-prod}

The outer product operation is crucial to the RIME, since it is used to characterize the coherency of two EMF or voltage vectors.  In tensor terms, the outer product can be defined (Sect.~\ref{sec:outer-prod}) in a completely abstract and coordinate-free way. By contrast, the definition usually employed in physics literature, and the RIME literature in particular, consists of a formal, mechanical manipulation of vectors (in the list-of-numbers sense). In particular, to derive the $4\times4$ formalism, \citet{ME1} \citep[see also Paper I,][Sect.~6.1]{RRIME1} used the Kronecker product form:

\[
\left( \begin{array}{c}x_1\\x_2\end{array} \right)\otimes
\left( \begin{array}{c}y_1\\y_2\end{array} \right) = 
\left( \begin{array}{c}x_1y_1\\x_1y_2\\x_2y_1\\x_2y_2\end{array} \right),
\]

while the Jones formalism is emerged \citep[Sect.~1]{ME4,RRIME1} by using the matrix product $\vec x \vec y^\herm$ instead. Note that while the former operation produces 4-vectors and the latter $2\times 2$ matrices, the two are isomorphic. It is important to establish whether
an outer product defined in this way is fully equivalent to that defined in tensor theory.

In fact, by defining the outer product as $\vec x \vec y^\herm$ in a \emph{specific coordinate system}, we're implicitly postulating a natural metric ($\mathbf{M}=\delta_{ij}$) in that coordinate system. This is of no consequence if only a single coordinate system is used, or if we restrict ourselves to unitary coordinate transformations, as is the case for transformations between $xy$ linearly polarized and $rl$ circularly polarized coordinates (such coordinate frames are called \emph{mutually unitary\/}). It is something to be kept in mind, however, if formulating the RIME in a coordinate-free way.

An outer product given by $\mathbf{V}=\vec x \vec y^\herm$ in a \emph{specific} coordinate system transforms as $\mathbf{A}^{-1}\mathbf{V}\mathbf{A}$ under change of coordinates, just like Jones matrices do. Alternatively, we may mechanically define an outer product-like operation as $\mathbf{W}=\vec x \vec y^\herm$ in \emph{all} coordinate systems, and this would then transform as $\mathbf{A}^{-1}\mathbf{V}[\mathbf{A}^{-1}]^\herm.$ The two definitions are only equivalent under unitary coordinate transformations! This is an easily overlooked point, most recently missed by the author of this paper: in Paper~I 
\citep[Sect.~6.3]{RRIME1}, only the second transform is given for coherency matrices, with no mention of the first. It may be somewhat academic in practice, since all applications of the RIME to date have restricted themselves to the mutually unitary $xy$ and $rl$ coordinate frames, but it may be relevant for future developments.

\subsubsection{Mixed-dimensionality tensors}
\label{sec:tensors-mixed-dim}

Under the strict definition, a tensor is associated with a single vector field $\mathbb{F}^N$, and so must be represented by an $N\times N\times...\times N$ array of scalars. In other words, all its indices must have the same range from $1$ to $N$.

Applied literature (and the present paper in particular) often makes use of mixed-dimensionality tensors (MDTs), i.e. arrays of numbers with different dimensions. In particular, Sect.~\ref{sec:tensor-rime} introduces the visibility MDT $\tens{V}^{pi}_{qj}$, with nominal dimensions of $N\times N\times 2\times2$. Such entities are not proper tensors in the strict definition, so we should formally establish to what extent they can be treated as such. The point is not entirely academic. Einstein summation (or any of the other operations discussed above) can be mechanically 
applied to arbitrary arrays of numbers, but the results are not guaranteed to be self-consistent under change of coordinates unless it can be formally established that they behave like tensors. Conversely, if we can formally establish that some operation yields a tensor of type $(n,m)$, then we know exactly how to transform it.

At first glance, MDTs seem to be significantly different from proper tensors. The difficulty lies in the fact that they seem to have two categories of indices. For example, $\tens{V}^{pi}_{qj}$  has ``2-indices'' (or ``3-indices'') $i,j$, associated with the vector space $\mathbb{C}^2$ or $\mathbb{C}^3$, with respect to which the MDT behaves like a proper tensor, and ``$N$-indices'' like $p$ and $q$, which only serve to ``bundle'' lower-ranked tensors together. $\tens{V}^{pi}_{qj}$ really behaves like a bundle of matrices (rank-2 tensors) rather than a proper rank-4 tensor. In particular, the coordinate transforms we normally consider involve the 2-indices only, and not the $N$-indices, so $\tens{V}^{pi}_{qj}$ really transforms like a (1-1)-type tensor. Fortunately, it turns out that MDTs \emph{can} be mapped to proper tensors in a mathematically rigorous way.

Let's assume we have a vector space like $\mathbb{C}^2$ (which we'll call the \emph{core space\/}), with a core metric of $g_{ij}$, and MDTs with a combination of 2-indices and $N$-indices. For illustration, consider the simpler case of the matrix $\tens{W}^i_p$, which (under the above terminology) is really a bundle of $N$ 2-vectors:

\[
  \mathbf{W} = \left( \begin{array}{cccc} w^1_1 & w^1_2 & \ldots & w^1_N \\ w^2_1 & w^2_2 & \ldots & w^2_N \end{array} \right )
\]

Let's formally map MDTs to conventional tensors over $\mathbb{C}^{2+N}$ space as follows:

\begin{equation}
\label{eq:mapping-mixed-dims}
  \hat\tens{V}^{pi}_{qj} = \left\{ 
    \begin{array}{lll} 
      \tens{V}^{(p-2)i}_{(q-2)j}&\;\;\;&\mbox{if}\;p>2,q>2,i\le 2,j\le 2\\
      0,&&\mbox{otherwise},
    \end{array}
    \right.
\end{equation}

This can be generalized to any mix of 2- and $N$-indices. We'll use the term \emph{2-restricted tensor} for any tensor over $\mathbb{C}^{2+N}$ whose components are null if any $N$-index is equal to 2 or less, or any 2-index is above 2. Note that this mapping from MDTs to 2-restricted tensors is isomorphic: every 2-restricted tensor over $\mathbb{C}^{2+N}$ has a unique MDT counterpart.

For the matrix $\tens{W}^i_p$, the mapping procedure effectively pads it out with nulls to make a $(N+2)\times(N+2)$ matrix:

\begin{equation}
\label{eq:padded-matrix}
  \hat\mathbf{W} = \left( 
    \begin{array}{cccccc}
      0 & 0 & w^1_1 & w^1_2 & \ldots & w^1_N \\ 
      0 & 0 & w^2_1 & w^2_2 & \ldots & w^2_N \\
      0 & 0 & 0 & 0 & \ldots & 0 \\
      \vdots & & & & & \vdots \\
      0 & 0 & 0 & 0 & \ldots & 0 \\
    \end{array}
  \right)
\end{equation}

In a sense, this procedure partitions the dimensions of the $\mathbb{C}^{2+N}$ space into the core dimensions (the first two), and the ``bundling'' dimensions (the other $N$). Formally, all the indices of $\hat\tens{V}$ and $\hat\tens{W}$ range from 1 to $N+2$, but 2-restricted tensors are constructed in such a way that components whose indices are ``out of range'' are all null. 

Now let's also map the coordinate transforms of the core space $\mathbb{C}^2$ onto a subset of the coordinate transforms of $\mathbb{C}^{2+N}$, using a transformation matrix of the form

\begin{equation}
\label{eq:restricted-xform}
  \mathbf{A} = \left( 
    \begin{array}{cccccc}
      a_1^1  & a_2^1 & 0       & 0 & \ldots & 0 \\ 
      a_1^2  & a_2^2 & 0       & 0 &        &   \\
      0      & 0     & 1       & 0 &        &   \\
      0      & 0     & 0       & 1 &        &   \\
      \vdots &       &         &   & \ddots & 0\\
      0      &       & \ldots  &   &     0  & 1  \\
    \end{array}
  \right)\;\;\;\mbox{(i.e. $a^i_j=\delta^i_j$ if $i$ or $j>2$}),
\end{equation}

where $\delta^i_j$ is the Kronecker delta. The $\hat\mathbf{W}$ matrix transforms as $\mathbf{A}^{-1}\hat\mathbf{W}\mathbf{A}$; it is easy to verify that it retains the same padded layout as Eq.~(\ref{eq:padded-matrix}) under such restricted coordinate transforms, and that the upper-right block of the padded matrix actually transforms to $\mathbf{A}_{(2)}^{-1} \mathbf{W}$, where $\mathbf{A}_{(2)}$ is the upper-left $2\times2$ corner of $\mathbf{A}$ (giving the original transform of $\mathbb{C}^2$). In other words, every vector $\vec w_j$ of the bundle transforms as $\mathbf{A}_{(2)}^{-1} \vec w_j$, exactly as vectors over the core space do!

This property generalizes to higher-rank tensors. For example, the 2-restricted tensor $\hat\tens{V}^{pi}_{qj}$ should formally transform as (see Eq.~(\ref{eq:tensor-coord-xform})):

\[
\hat\tens{V'}^{pi}_{qj} = \tilde a^p_\sigma \tilde a^i_\alpha a^\tau_q a^\beta_j \hat\tens{V}^{\sigma\alpha}_{\tau\beta},
\]

However, for $p\le2$ we have $\hat\tens{V}^{\sigma\alpha}_{\tau\beta}=0$ by definition, while for $p>2$, $\tilde a^p_\sigma=1$ for $p=\sigma$, and is null otherwise. Therefore, only the $p=\sigma$ (and, similarly, $q=\tau$) terms contribute to the sum above. Thus,

\[
\hat\tens{V'}^{pi}_{qj} = \tilde a^i_\alpha a^\beta_j \hat\tens{V}^{p\alpha}_{q\beta},
\]
so our nominally (2,2)-type tensor $\hat\tens{V}^{pi}_{qj}$ behaves exactly like a (1,1)-type tensor under any coordinate transform given by 
Eq.~(\ref{eq:restricted-xform}). The $p$ and $q$ indices can be called \emph{invariant} (as opposed to co- or contravariant). In general, any 2-restricted $(n,m)$-type tensor having $(n',m')$ 2-indices always behaves as an $(n',m')$-type tensor under such coordinate transforms.

For the same reason, we can relax the rules of Einstein summation to allow \emph{repeated} invariant indices. For example, $z^i = x^i y^i$ is not a valid tensor (one index, but transforms doubly-contravariantly!), but $\tens{Z}^i_p = \tens{X}^j_p \tens{Y}^i_{jp}$ is a perfectly valid tensor, since an extra invariant index $p$ does not change how the components transform. 

Furthermore, it is easy to see that a 2-restricted tensor remains 2-restricted under any coordinate transform of the core vector space, so the property of being 2-restricted is, in a sense, intrinsic. Any product of 2-restricted tensors is also 2-restricted. In other words, 2-restricted tensors form a closed subset under coordinate transforms of the core vector space, and under all product operations. To complete the picture, we can define a metric in $\mathbb{C}^{2+N}$ by using the core metric for the core dimensions, and the natural metric for the ``bundling'' dimensions, so that tensor conjugation is expressed in terms of the core metric only:

\[
  \bar{\tens{V}}^{pi}_{qj} = g_{i\alpha}g^{j\beta} \overline{ {\tens{V}}_{p\beta}^{q\alpha} }.
\]

To summarize, we have formally established that MDTs with a core vector space of $\mathbb{C}^M$ and a second\footnote{Additional dimensionalities may be ``mixed in'' in the same way.} dimensionality of $N$ can be isomorphically mapped onto the set of $M$-restricted tensors over $\mathbb{C}^{M+N}$. Under coordinate transforms of the core vector space, such tensors behave co- and contravariantly with respect to the $M$-indices, and invariantly with respect to the $N$-indices. We are therefore entitled to treat MDTs as proper tensors for the purposes of this paper.

\bibliographystyle{aa}

\bibliography{16764}

\end{document}